\documentclass[twocolumn,english,superscriptaddress]{revtex4}
\usepackage[T1]{fontenc}
\usepackage[utf8]{inputenc}
\usepackage{bm}
\usepackage{amsmath}
\usepackage{amssymb}
\usepackage{graphicx}
\usepackage{wasysym}
\usepackage[caption=false]{subfig}

\makeatletter

\newcommand{\bra}[1]{{\left\langle #1 \right|}}
\newcommand{\ket}[1]{{\left| #1 \right\rangle}}

\newcommand{\refeq}[1]{{(\ref{#1})}}
\@ifundefined{textcolor}{}
{%
 \definecolor{BLACK}{gray}{0}
 \definecolor{WHITE}{gray}{1}
 \definecolor{RED}{rgb}{1,0,0}
 \definecolor{GREEN}{rgb}{0,1,0}
 \definecolor{BLUE}{rgb}{0,0,1}
 \definecolor{CYAN}{cmyk}{1,0,0,0}
 \definecolor{MAGENTA}{cmyk}{0,1,0,0}
 \definecolor{YELLOW}{cmyk}{0,0,1,0}
}

\usepackage[colorlinks,linkcolor=blue,anchorcolor=blue,citecolor=blue]{hyperref}

\makeatother

\usepackage{babel}
\begin{document}

\title{Charge density wave and charge pump of interacting fermions in circularly
shaken hexagonal optical lattices}

\author{Tao Qin}

\affiliation{Institut für Theoretische Physik, Goethe-Universität, 60438 Frankfurt/Main,
Germany}

\author{Alexander Schnell}

\affiliation{Max-Planck-Institut für Physik komplexer Systeme, Nöthnitzer Straße
38, 01187 Dresden, Germany}

\author{Klaus Sengstock }

\affiliation{Institut für Laserphysik (ILP), Universität Hamburg, Luruper Chaussee
149, 22761 Hamburg, Germany}

\affiliation{Hamburg Centre for Ultrafast Imaging, Luruper Chaussee 149, 22761
Hamburg, Germany}

\affiliation{Zentrum für Optische Quantentechnologien (ZOQ), Universität Hamburg,
Luruper Chaussee 149, 22761 Hamburg, Germany}

\author{Christof Weitenberg}

\affiliation{Institut für Laserphysik (ILP), Universität Hamburg, Luruper Chaussee
149, 22761 Hamburg, Germany}

\affiliation{Hamburg Centre for Ultrafast Imaging, Luruper Chaussee 149, 22761
Hamburg, Germany}

\author{André Eckardt}

\affiliation{Max-Planck-Institut für Physik komplexer Systeme, Nöthnitzer Straße
38, 01187 Dresden, Germany}

\author{Walter Hofstetter}

\affiliation{Institut für Theoretische Physik, Goethe-Universität, 60438 Frankfurt/Main,
Germany}
\begin{abstract}
We analyze strong correlation effects and topological properties
of interacting fermions with a Falicov-Kimball type interaction
in circularly shaken hexagonal optical lattices, which can be effectively
described by the Haldane-Falicov-Kimball model, using 
the real-space Floquet dynamical mean-field theory (DMFT). 
The Haldane model, a paradigmatic model of the Chern insulator, is experimentally relevant, because it has been realized using circularly shaken hexagonal optical lattices.
We show that in the presence of staggering a charge density
wave emerges, which is affected by interactions and resonant tunneling.
We demonstrate that interactions smear out the edge states by introducing a
finite life time of quasiparticles. Even though a general method
for calculating the topological invariant of a nonequilibrium steady state
is lacking, we extract the topological invariant using a Laughlin
charge pump set-up. We find and attribute to the dissipations into the bath connected to every lattice site, which is intrinsic to real-space Floquet DMFT methods, that
the pumped charge is not an integer even for the non-interacting case at very low reservoir temperatures.
Furthermore, using the rate equation based on  the Floquet-Born-Markov approximation,
we calculate the charge pump from the rate equations for the non-interacting
case to identify the role of the spectral properties of the bath. Starting from this approach we propose an experimental protocol for measuring quantized charge pumping. 
\end{abstract}
\maketitle

\section{Introduction}

Time periodically driven ultracold atoms in optical lattices are a
versatile and powerful platform to simulate models with non-trivial
topological properties~\cite{Eckardt2017aqg,Goldman:2016aa}. Two
paradigmatic models, the Hofstadter model and the Haldane model have
been realized with Raman laser assisted tunneling~\cite{Bloch2013,Ketterle2013,Aidelsburger:2015aa}
and circularly shaken hexagonal lattices~\cite{Jotzu:2014fk,Flaschner:2016ab,Flaschner:2017aa},
respectively. Different techniques have been developed in setups with
ultracold atoms  in optical lattices to detect topological properties.
Using a drift measurement, 
the topology of the lowest band of the Hofstadter model was determined~\cite{Aidelsburger:2015aa}.
By measuring the
shift of atom clouds in a one-dimensional superlattice, the Thouless
charge pump was realized in bosonic~\cite{Lu2016gpb, Lohse:2016aa} and fermionic~\cite{Nakajima:2016aa}
systems. A two-dimensional version of the topological charge pump
was demonstrated by mapping a four-dimensional quantum Hall system
to a two-dimensional square superlattice using dimensional reduction~\cite{Lohse:2018aa}.
Using the tomographic technique that was proposed in Ref.~\cite{Hauke2014},
the Berry curvature of the Haldane model was mapped out in momentum
space~\cite{Flaschner:2016ab}. Also, dynamical vortices due to quenching
into the Floquet Hamiltonian were observed~\cite{Flaschner:2017aa},
which are a non-equilibrium signature of topology. The trajectories
of the dynamical vortices in momentum space were used to determine
the linking number~\cite{Tarnowski:2017aa}, which can be directly
related to the Chern number of the Hamiltonian after a quench~\cite{Wang2017smt}. So far, one can understand most experimental achievements~\cite{Jotzu:2014fk,Flaschner:2016ab,Flaschner:2017aa,Tarnowski:2017aa} with non-interacting effective Hamiltonians~\cite{bukov2015}.

Introducing two-particle interactions into a time-periodically driven
system is a highly non-trivial problem from the point of view of both
experiment and theory. In experiments one needs to overcome the problem
of heating. It was shown that an interacting time-periodically driven
closed system will heat up to a trivial state with infinite
temperature~\cite{DAlessio2014,Lazarides2014esg}, with only a few exceptions like many-body localized systems~\cite{Ponte2015mbl,Lazarides2015fmb},
integrable systems~\cite{Lazarides2014pti} and the prethermalization plateau~\cite{Bukov2015pm,Weidinger:2017aa}. 
Multi-photon interband heating has been observed in a shaken 1D optical lattice~\cite{Weinberg2015mie}. Resonant tunneling, which happens when interactions are integer multiples of the driving frequency and magnetic correlations have been measured for strongly correlated fermions in hexagonal optical lattices with periodic driving in one direction~\cite{Gorg:2018aa}.
Floquet evaporative
cooling was shown to reduce heating for interacting bosons in a one-dimensional optical
lattice~\cite{Reitter2017idh}. However, there are no 
artificial gauge fields in these setups. Further efforts are needed
to go into the interacting regime and realize an interacting system with
artificial gauge fields. Theoretically, in the high-frequency limit,
where the time periodically driven system is supposed to be in a prethermalized
regime~\cite{Bukov2015pm}, the system can be described by an effective
Hamiltonian in high-frequency approximation~\cite{Goldman:2014ab,bukov2015,Eckardt2015hfa}. With interactions turned on, the interacting
Haldane model can be studied with the static mean-field approximation~\cite{Zheng2015mod,Plekhanov:2016aa} and exact diagonalization~\cite{Anisimovas2015rrs}.
The possible drawback of the effective Hamiltonian approach
is that it cannot describe the non-equilibrium properties of the system.
In the strongly correlated regime, one can obtain the effective low-energy
Hamiltonian in the limit of large interaction using a high-frequency
expansion, which is equivalent to a Schrieffer-Wolff transformation~\cite{Bukov2016sf}.
With it one can qualitatively analyze the properties of near-resonant
and resonant tunneling. However, to solve the low energy effective
Hamiltonian is still a very non-trivial many-body problem. 
Numerical tools such as quantum Monte-Carlo and density matrix renormalization method need to be adopted to solve it. 

Based on the experimental progress~\cite{Flaschner:2016ab,Flaschner:2017aa,Tarnowski:2017aa},
we investigate non-equilibrium steady states (NESS) of fermions with
Falicov-Kimball type interactions in a circularly shaken hexagonal
optical lattice in a non-perturbative way using the method of real-space Floquet
dynamical mean field theory (DMFT)~\cite{Freericks2006,Aoki2008,Freericks2008,Freericks2008nds,Aoki2014},
which can deal with driving, interactions and dissipation on equal
footing. We study the strong correlation effects and topological
properties of the system. We investigate the charge density wave (CDW) induced
by the staggered potential as a function of increasing interactions. To study 
topological properties, we use a Laughlin charge pump setup~\cite{Laughlin1981},
where a flux is inserted in the direction of the axis of a cylinder
geometry. We observe how the edge states are smeared out by interactions.
Furthermore, we calculate the charge pump due to insertion of
flux quanta for different interactions. The dissipation into the bath
makes the pumped charge non-integer. In addition, we study
the role of dissipation for the non-interacting case in the presence of a heat bath using rate equations based on the Floquet-Born-Markov approximation. We start from an initial state which is close to equilibrium, and ramp up the flux adiabatically to calculate the pumped charge. By comparison with the equilibrium case we believe our procedure is experimentally practical. 

We include a bath in all our calculations. While the bath prevents serious heating of the driven system, it causes dissipation which smears the integer charge pump, at least within our theoretical approaches for dealing with the bath. Whether it is possible to recover the integer charge pump by bath engineering will be a future direction.

The manuscript is organized as follows. In Sec.~\ref{sec:MM}, we
present the model and methods used in our calculations. We outline the real-space Floquet DMFT method for the interacting and driven system, and the rate equation for the non-interacting case. In Sec.~\ref{sec:RD},
we present our results on the charge density wave and charge pump for the interacting system. For the non-interacting case, we show calculations from rate equations. We conclude in Sec.~\ref{sec:Cc}.

\section{\label{sec:MM}Model and methods}

\subsection{The model}
We start with a model for fermions in circularly shaken hexagonal
optical lattices which can describe the experimental setups described in Refs.~\cite{Flaschner:2016ab,Flaschner:2017aa,Tarnowski:2017aa,Sohal:2016aa}
\begin{equation}
H_{0}=-J\sum_{\left\langle ll^{\prime}\right\rangle }c_{l^{\prime}}^{\dagger}c_{l}+\sum_{l}\nu_{l}\left(t\right)n_{l}+\alpha\left(\Omega+\Delta\right)\sum_{l}\lambda_{l}n_{l},\label{eq:H0t}
\end{equation}
where $l$ and $l^{\prime}$ label lattices sites, and $n_{l}=c_{l}^{\dagger}c_{l}$.
$\Omega$ is the driving frequency and $\Delta$ is the detuning between the driving frequency and the AB-offset in the static lattice.
$\lambda_{\mathrm A\left(\mathrm B\right)}=0\left(1\right)$ for A and B sites in
the unit-cell. $\alpha=\pm1$ which is the sign of the staggered potential. We consider a near-resonant driving, which reestablishes resonant tunneling between A and B sites.
The driving term is given by $\nu_{l}\left(t\right)=-\bm{r}_{l}\cdot\bm{F}\left(t\right)=-\bm{r}_{l}\cdot F\left[\cos\left(\Omega t\right)\hat{\bm{e}}_{x}+\tau\sin\left(\Omega t\right)\hat{\bm{e}}_{y}\right]$,
where $\tau=\pm1$ corresponds to counter-clockwise (clockwise) shaking.
In the following, we choose $\alpha=1$ and $\tau=1$.  For the Schr\"odinger
equation of the system $i\frac{d\left|\psi\right\rangle }{dt}=H_{0}\left|\psi\right\rangle $,
with the unitary transformation $\left|\psi\right\rangle =\mathcal{U}\left|\tilde{\psi}\right\rangle $
and $\mathcal{U}=e^{i\sum_{l}\tilde{\chi}_{l}\left(t\right)n_{l}}$, 
where 
\begin{align}
\tilde{\chi}_{l}\left(t\right)&=-\int_{0}^{t}\tilde{\nu}_{l}\left(t\right)dt+\frac{1}{\mathcal{T}}\int_{0}^{\mathcal{T}}dt^{\prime\prime}\int_{0}^{t^{\prime\prime}}dt^{\prime}\tilde{\nu}_{l}\left(t^{\prime}\right),\\
\tilde{\nu}_{l}\left(t\right)&=\nu_{l}\left(t\right)+\alpha\Omega\lambda_{l},
\end{align}
and $\mathcal{T}=\frac{2\pi}{\Omega}$, we have $\tilde{H}_{0}=\mathcal{U}^{\dagger}H_{0}\mathcal{U}-i\mathcal{U}^{\dagger}\frac{d}{dt}\mathcal{U}$.
Therefore, 
\begin{equation}
\tilde{H}_{0}(t)=-J\sum_{\left\langle ll^{\prime}\right\rangle }e^{i\tilde{\theta}_{l^{\prime}l}\left(t\right)}c_{l^{\prime}}^{\dagger}c_{l}+\alpha\Delta\sum_{l}\lambda_{l}n_{l}\label{eq:H0tilde}
\end{equation}
where $\tilde{\theta}_{l^{\prime}l}\left(t\right)=\frac{K}{\Omega}\sin\left(\Omega t-\tau\phi_{l^{\prime}l}\right)+\alpha\epsilon_{l^{\prime}l}\Omega t-\alpha\epsilon_{l^{\prime}l}\pi$.
$\phi_{l^{\prime}l}$ is defined by $\bm{r}_{l^{\prime}}-\bm{r}_{l}=\cos\left(\phi_{l^{\prime}l}\right)\hat{\bm{e}}_{x}+\sin\left(\phi_{l^{\prime}l}\right)\hat{\bm{e}}_{y}$
for nearest neighbors. 

We consider a Falicov-Kimball interaction, where the mobile atoms
interact with localized atoms:
\begin{equation}
H_{\mathrm{int}}=U\sum_{l}c_{l}^{\dagger}c_{l}f_{l}^{\dagger}f_{l}.
\end{equation}
$U$ is the interaction strength. $f_{l}$ ($f_{l}^{\dagger}$) is the
annihilation (creation) operator for localized atoms. $f$-atoms work as an annealed disorder for the mobile atoms. The interaction term $H_{\mathrm{int}}$ is not affected by the driving term in Eq.~\eqref{eq:H0t}, because it commutes with the driving term.

\subsection{Real-space Floquet DMFT}
We outline the real-space Floquet DMFT method that we adopt
to deal with the Falicov-Kimball interaction~\cite{Freericks2006,Aoki2008,Tsuji2010,Qin2017sft}.
It is a method to study the NESS in an inhomogeneous system. To reach the
NESS, every lattice site is coupled to a bath. We use a free-fermion
bath in our implementation~\cite{Aoki2009,Aoki2014}. The full Green's
function of the lattice system satisfies Dyson's equation, 
\begin{align}
\left(\hat{G}^{-1}\right)_{ll^{\prime},mn}\left(\omega\right)= & \left(\hat{G}_{0}^{-1}\right)_{ll^{\prime},mn}\left(\omega\right)-\hat{\Sigma}_{l,mn}\left(\omega\right)\delta_{ll^{\prime}}\nonumber \\
 & -\hat{\Sigma}_{\mathrm{bath},l,mn}\left(\omega\right)\delta_{ll^{\prime}}
\end{align}
where every part is defined on the Keldysh contour~\cite{rammer2007quantum}
and in Floquet space 
\begin{equation}
\hat{G}\left(\omega\right)=\left(\begin{array}{cc}
G^{R} & G^{K}\\
0 & G^{A}
\end{array}\right)\left(\omega\right).
\end{equation} $\omega\in\left[-\frac{\Omega}{2},\frac{\Omega}{2}\right)$ is in
the first Brillouin zone of $\Omega$. For the non-interacting part we have
\begin{equation}
G_{0ll^{\prime},mn}^{R-1}\left(\omega\right)=\left(\omega-\mu+n\Omega+i0^{+}\right)\delta_{mn}\delta_{ll^{\prime}}-\tilde{H}_{0ll^{\prime},mn},
\end{equation}
where $\mu$ is the chemical potential, $\tilde{H}_{0ll^{\prime},mn}=\frac{1}{\mathcal{T}}\int_{0}^{\mathcal{T}}dt\tilde{H}_{0ll^{\prime}}\left(t\right)e^{i\left(m-n\right)\Omega t}
$ with $\tilde{H}_{0,ll^{\prime}}\left(t\right)$ defined $\tilde{H}_{0}\left(t\right)=\sum_{\left\langle ll^{\prime}\right\rangle }c_{l^{\prime}}^{\dagger}\tilde{H}_{0,ll^{\prime}}\left(t\right)c_{l}$. The Floquet indices $m$ and $n$ are integers, and $-\infty<m,n<\infty$. In our calculations, by choosing the matrix of finite size for the Green's function in the Floquet space, we include several Floquet bands. When the driving frequency is large, one can achieve a convergent calculation with a small Floquet matrix size~\cite{Aoki2008}. The Hamiltonian $\tilde{H}_0(t)$ enters the calculation through its form in the Floquet space. In the high driving frequency limit, it can be related to the effective Haldane Hamiltonian by $\tilde{H}_{0,\mathrm{eff}}=\tilde{H}_{0,0}+\sum_{m=1}^{\infty}\frac{1}{m\Omega}\left[\tilde{H}_{0,m},\,\tilde{H}_{0,-m}\right]+\cdots$,  where we have defined $\tilde{H}_{0,m-n}\equiv\tilde{H}_{0,m,n}$~\cite{Eckardt2015hfa,Tarnowski:2017aa} and omit the spatial indices.

Furthermore, $G_{0}^{A}\left(\omega\right)=G_{0}^{R\dagger}\left(\omega\right)$,
as well as $\left(G_{0}^{-1}\left(\omega\right)\right)^{K}=0$~\cite{Aoki2009,Aoki2014}. The coupling between the system (s) and the bath (b) is~\cite{Aoki2014},
$H_{s-b}=\sum_{i,p}V_{p}\left(c_{i}^{\dagger}b_{i,p}+b_{i,p}^{\dagger}c_{i}\right)$
where $b_{i,p}$ ($b_{i,p}^{\dagger}$) is the fermion annihilation (creation) operator for the
bath. 
$\Sigma_{\mathrm{bath},l,mn}\left(\omega\right)$ is the correction to the self-energy on site $l$
due to dissipation to the bath 
\begin{equation}
\Sigma_{\mathrm{bath},l,mn}\left(\omega\right)=\left(\begin{array}{cc}
-i\Gamma\delta_{mn} & -2i\Gamma F_{n}\left(\omega\right)\delta_{mn}\\
0 & i\Gamma\delta_{mn}
\end{array}\right),
\end{equation} 
assuming that the density of states (DOS) of the bath is constant. 
 $\Gamma$ is a phenomenological dissipation rate to the bath, and $F_{n}\left(\omega\right)=\tanh\frac{\omega+n\Omega}{k_{B}T}$ 
where $T$ is the temperature of the bath~\cite{Aoki2014}. $\Sigma_{l,mn}\left(\omega\right)$ is
the lattice self-energy due to two-particle interactions and is obtained
from the impurity solver for every lattice site $l$: \begin{equation}G_{l}\left(\omega\right)=w_{0}\mathcal{G}_{0,l}\left(\omega\right)+w_{1}\left[\mathcal{G}_{0,l}^{-1}\left(\omega\right)-U\right]^{-1}\label{solver}\end{equation}
where $w_{1}$ is the probability of one site being occupied by immobile
atoms and $w_{0}=1-w_{1}$. Equation~\eqref{solver} is the exact solution for the Falicov-Kimball model of infinite dimensions. We refer the reader to Refs.~\cite{Brandt:1989aa, Eckstein2008nss,Supple2018} for the technical details of the solution.  In the following, we focus on the case
of half filling, for which $w_{1}=\frac{1}{2}$ and $w_{0}=\frac{1}{2}$.
The self-consistent loop is closed by \begin{equation}\mathcal{G}_{0,l}^{-1}\left(\omega\right)=G_{l}^{-1}\left(\omega\right)+\Sigma_{l}\left(\omega\right).\end{equation}

\subsection{Rate Equations in presence of heat bath}
Here we present a method of studying the NESS using rate equations for the non-interacting gas. 
Using this approach we will investigate the impact of the spectral properties of the bath on the non-equilibrium steady state of the system and the quantization of charge pumping. 

In order to  access situations where the particle number~$N$ of fermions 
in the system is conserved and there is only heat exchange with a thermal environment, we 
here present an alternative treatment using rate equations.
This method only applies to the noninteracting Fermi gas,
where $U=0$.

Here, the total Hamiltonian reads
\begin{equation}
	H(t)=\tilde{H}_0(t) + \sqrt{\Gamma} v \sum_\alpha \kappa_\alpha (b^\dagger_{\alpha} + b_{\alpha}) + \sum_\alpha \omega_\alpha b^\dagger_{\alpha} b_{\alpha},
\end{equation}
where the bath is modeled by a collection of harmonic oscillators $b_\alpha$, corresponding frequencies $\omega_\alpha$ and 
dimensionless coupling constants $\kappa_\alpha$, and some system coupling operator $v$.
Note that we have separated the strength $\sqrt{\Gamma}$ of the system--bath coupling
from the coefficients $\kappa_\alpha$. It turns out that the magnitude of the dissipation rate is given 
essentially by $\Gamma$.

In the weak system--bath coupling limit, $\Gamma \to 0$,
 we may perform the usual Born-Markov \cite{BreuerPetruccione} and the full rotating wave approximation 
\cite{Kohler98, Hone2009, cohen1994atoms} in which we average over the long relaxation time scales $\propto {1/\Gamma}$
(rather than just one period of the driving). For a single fermion, $N=1$, one then finds that the reduced system density matrix is asymptotically diagonal 
in the Floquet states $\ket{a(t)}$, i.e.~$\varrho(t) = \sum_{a} p_a(t) \ket{a(t)}\bra{a(t)}$, and the asymptotic dynamics is governed
by a Pauli rate equation, 
\begin{equation}
	\partial_t p_a(t) = \sum_b \left[R_{ab} p_b(t) - R_{ba} p_a(t)\right],
\end{equation}
that describes the transfer between populations $p_a(t)$ of the Floquet states.
This happens at a rate 
\begin{equation}
	R_{ab} = {2\pi} \Gamma \sum_{m\in \mathbb{Z}} \vert v_{ab}^{(m)} \vert^2 
	g(\varepsilon_a-\varepsilon_b-m\Omega),
	\label{eq:rate-sp}
\end{equation}
involving the quasienergy $\varepsilon_a$ of Floquet state $a$ and the $m$-th component of the Fourier transform of the coupling,
\begin{equation}
	v_{ab}^{(m)} = \frac{1}{\mathcal{T}} \int_0^{\mathcal{T}} \mathrm{d}t \bra{a(t)} v \ket{b(t)}  \mathrm{e}^{\mathrm{i}m\Omega t}.
\end{equation}
It also enters the bath-correlation function $g$ that reads for the phonon bath 
\begin{equation}
	g(E) =  \left\lbrace \begin{array}{cc}J(E) n_B(E),& E>0, \\
		J(-E) (1+n_B(-E)), & E<0,
	\end{array}\right.
\end{equation}
with the occupation function $n_B(E) = {1}/({\mathrm{e}^{E/T}-1} )$ and
the spectral density of the bath $J(E) = \sum_\alpha \kappa_\alpha^2 \delta(E-\omega_\alpha)$, for $E \geq 0$.
Typical baths with a continuum of modes $\alpha$ obey 
\begin{equation}
	J(E) \propto E^d {\mathrm{e}^{-E/E_c}},
\end{equation}
where the exponent $d$ controls the low-frequency behaviour of $J(E)$. Here $d=1$ denotes the ohmic case 
and $d < 1$ ($d>1$) is sub-(super-)ohmic. The high-frequency cutoff parameter $E_c$ basically is set by the
 correlation time $\tau_B \propto {1}/{E_c}$  of the bath \cite{BreuerPetruccione}.
 In order to be consistent with the Markov approximation,
 this time $\tau_B$ must be small when compared to the typical time scale of relaxation $\tau_R \propto 1/\Gamma$, 
 which is always valid in the weak coupling limit $\Gamma \rightarrow 0$ that we aim at.

The non-interacting Fermi gas may be considered in the same framework, 
however, one has to additionally implement quantum statistics.  This leads to
the many particle version of the Pauli rate equation \cite{Vorberg2013gbe},
\begin{equation}
	\partial_t \langle n_{a} \rangle =  \sum_{b} R_{ab} (1- \langle n_{a} \rangle) \langle n_{b} \rangle - R_{ba} (1-\langle  n_{b}  \rangle ) \langle n_{a} \rangle,
	\label{eq:kinetic}
\end{equation}
where $\langle n_{a} \rangle$ is the mean occupation of the Floquet state $a$ and where we have applied
the mean field approximation $\langle n_an_b\rangle\approx \langle n_a\rangle\langle n_b\rangle$ discussed in Ref.~\cite{Vorberg2015nss}.
The nonequilibrium steady state is found by solving for steady occupations,  $\partial_t \langle n_{a} \rangle =0$.

\section{\label{sec:RD}Results and discussions}

\subsection{Charge density wave\label{sec_cdw}}

\begin{figure}[h]
\includegraphics[scale=0.75]{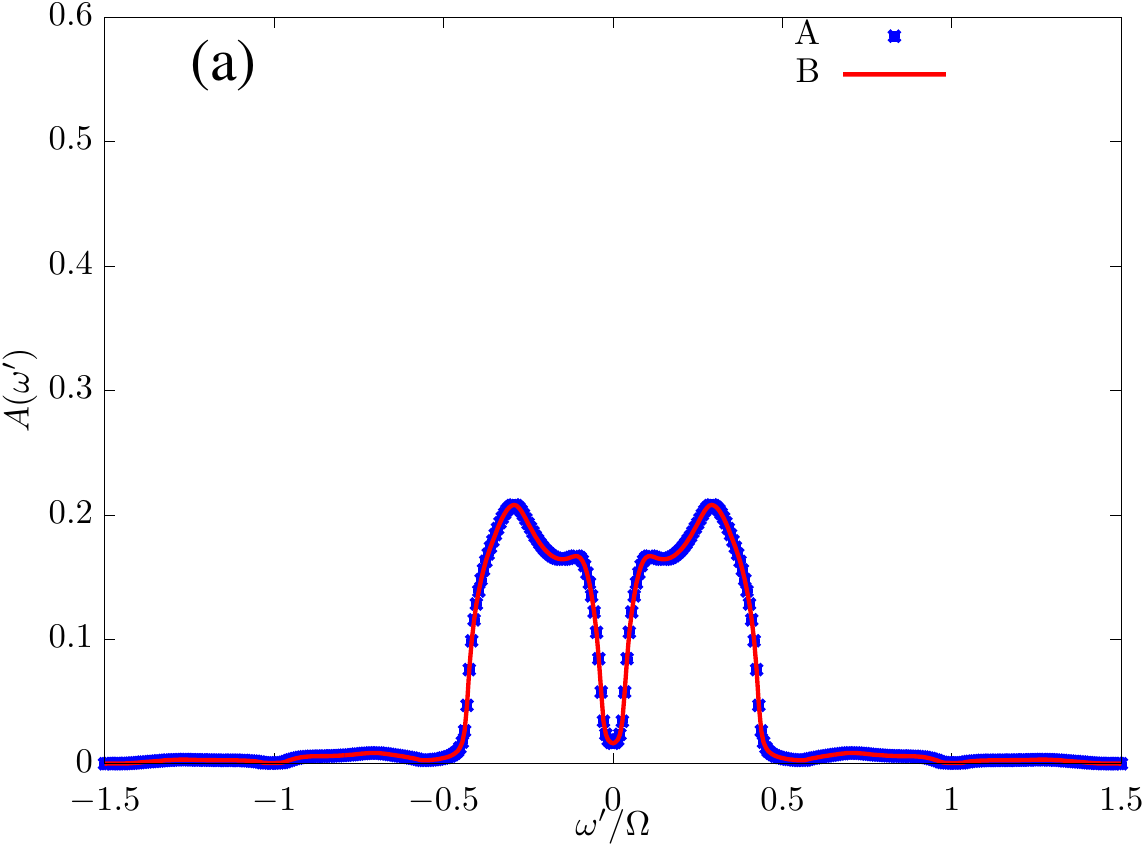}

\includegraphics[scale=0.75]{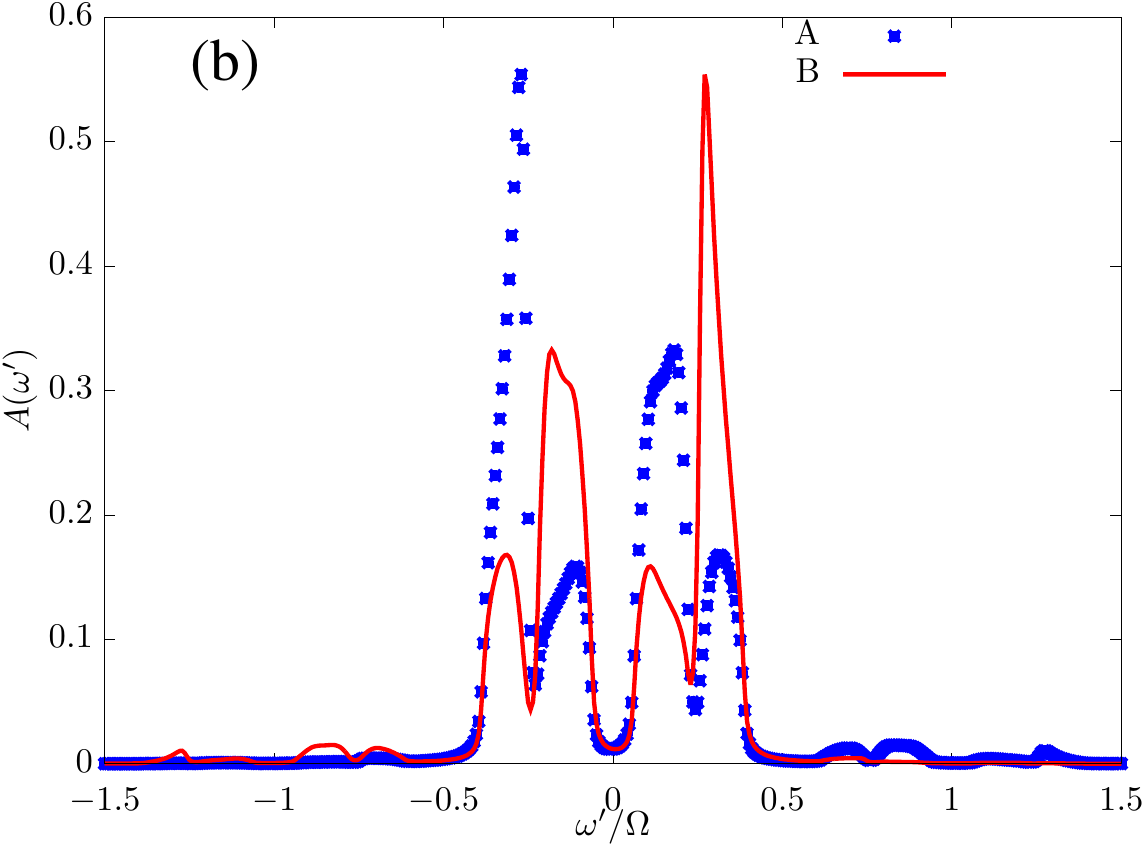}

\caption{\label{fig:Spectral}Spectral functions $A\left(\omega^\prime\right)=-\frac{1}{\pi}\mathrm{Im}G^R_{nn}\left(\omega\right)$ with $\omega^\prime=\omega+n\Omega$ for the case without staggered
potential in (a), and the case with staggered potential $\Omega\sum_{l}\lambda_{l}n_{l}$ in (b). The unit of the spectral function is [TL$^{-2}$].
For both panels, the driving frequency $\Omega=7J$, $\frac{K}{\Omega}=1.28$, and $\Delta=0$. Bath
parameters are $\Gamma=T=0.05J$. }
\end{figure}

In this section, we focus on the effects of interactions and resonant
tunneling on the charge density wave (CDW) induced by the staggered potential. We consider a two-dimensional hexagonal
optical lattice with $9\times9$ unit cells and periodic boundary
conditions in both $x$ and $y$ directions. The charge densities are different on A and B sites, with the definition of the charge
density on site A (B) $N_{i}^{c}=\frac{1}{2\pi}\int_{-\Omega/2}^{\Omega/2}d\omega\sum_{n}\mathrm{Im}G_{nn,i}^{<}\left(\omega\right)$,
where $i=\mathrm{A}\,(\mathrm{B})$, $n$ is the Floquet index for the Floquet Green's function $G^<_{A\left(B\right)}\left(\omega\right)$
on site A (B), and $G_{nn,i}^{<}\left(\omega\right)=\frac{1}{2}(G_{nn,i}^{K}\left(\omega\right)-G_{nn,i}^{R}\left(\omega\right)+G_{nn,i}^{A}\left(\omega\right))$. Because of the periodic boundary conditions, there are
only two different sites in the hexagonal optical lattice.  Our study
is different from those in Ref.~\cite{Chen2003cdw,Matveev2008odp,Nguyen2013cdp},
where the CDW order parameter is defined as $\Delta_{\mathrm{CDW}}^{f}=N_{\mathrm{A}}^{f}-N_{\mathrm{B}}^{f}$
because a finite density difference of localized $f$-atoms is needed
to spontaneously break the symmetry between sublattice sites A and
B. However, in our case of the real space implementation of DMFT calculations,
we choose $w_{1}=\frac{1}{2}$ for all sites. 
We nevertheless have a CDW also for $\Delta=0$. Namely, for  an integer $\alpha$, i.e.\ in the presence of the staggered potential $\alpha\Omega\sum_l\lambda_ln_l$, will cause an effective energy offset between A and B sites (appearing in the second-order high-frequency expansion of the effective Hamiltonian~\cite{Tarnowski:2017aa}). It results from virtual second-order processes where a particle tunnels from an A (B) site to a neighboring B (A) site and back.
In Fig.~\ref{fig:Spectral},
we show a comparison calculation to prove this. We can see a perfect
symmetry of the spectral functions for both A and B sites for the case without
staggered potential (panel (a)), in contrast to a broken symmetry for the case when
the staggered potential is present (panel (b)). 

\begin{figure}[h]
\includegraphics[scale=0.75]{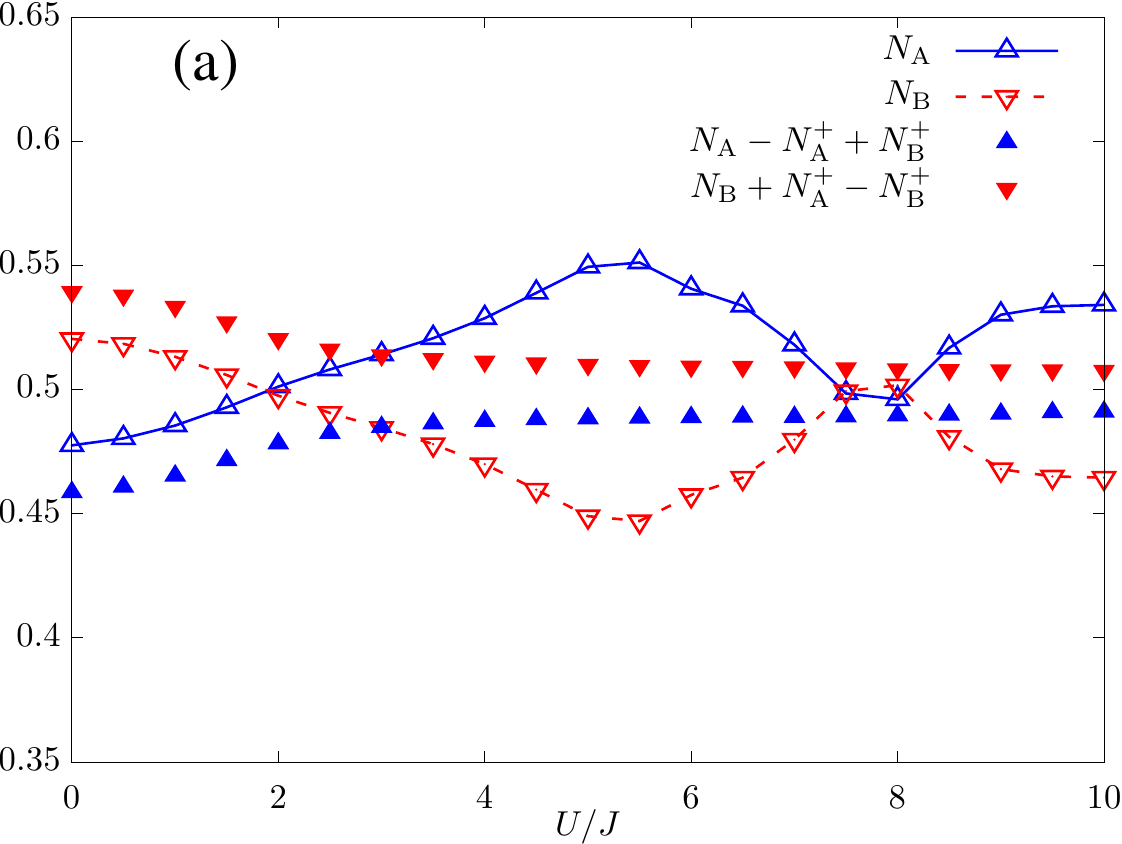}

\includegraphics[scale=0.75]{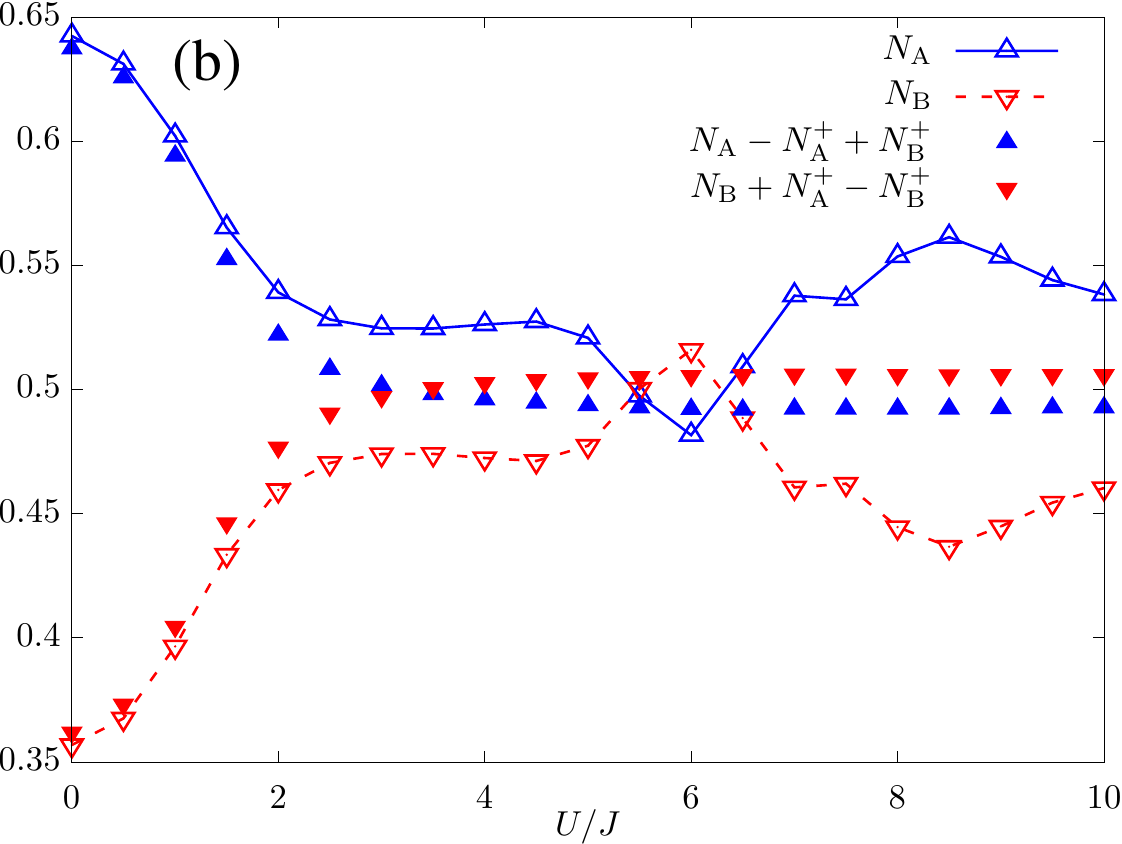}

\caption{\label{fig:cdw}Charge density wave (CDW) for (a) $\Delta=-0.6J$ and (b) $\Delta=0$ versus interactions $U$
for a two-dimensional hexagonal optical lattice with $9\times9$ unit
cells and periodic boundary conditions in both $x$ and $y$ directions. For both panels, the driving frequency $\Omega=7J$ and $ \frac{K}{\Omega}=1.28$.
Bath parameters are $\Gamma=T=0.05J$. $N_{\mathrm{A}\left(\mathrm{B}\right)}^{+}$ is
shown in Fig.~\ref{fig:cdw>0}. }
\end{figure}

\begin{figure}[h]
\includegraphics[scale=0.75]{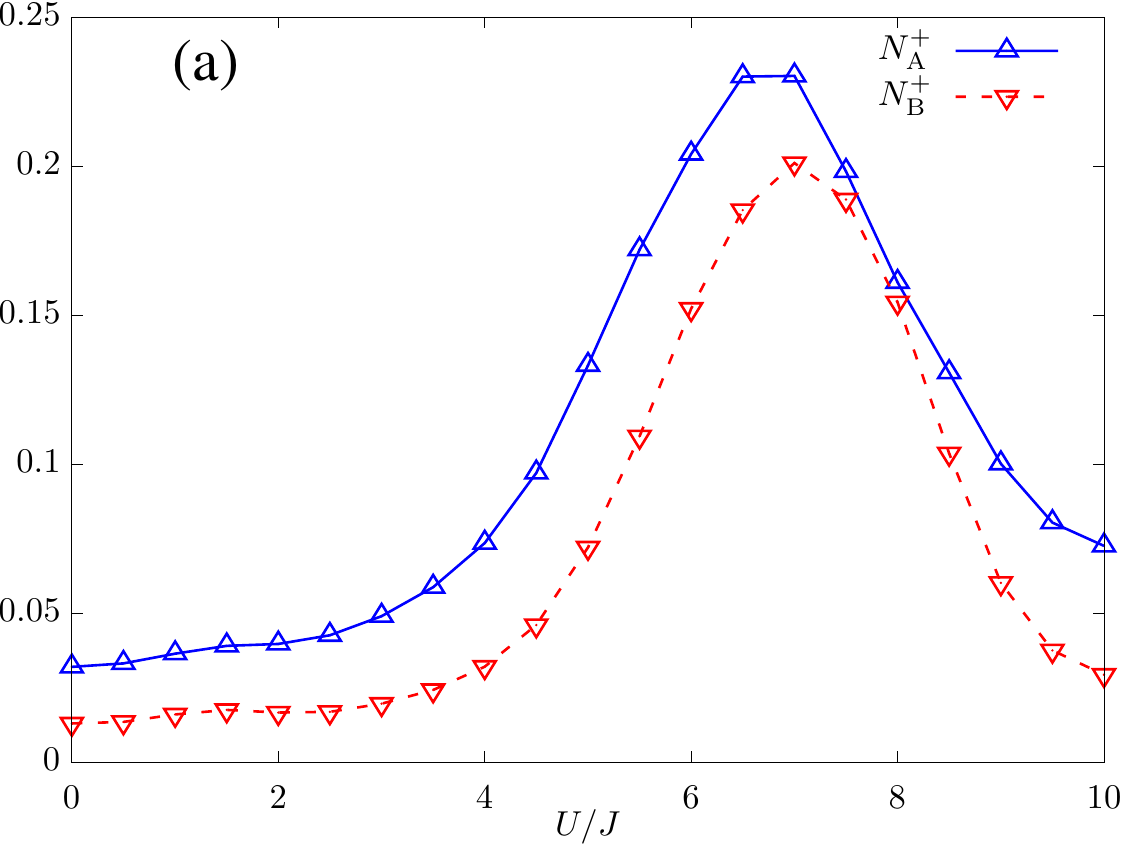}

\includegraphics[scale=0.75]{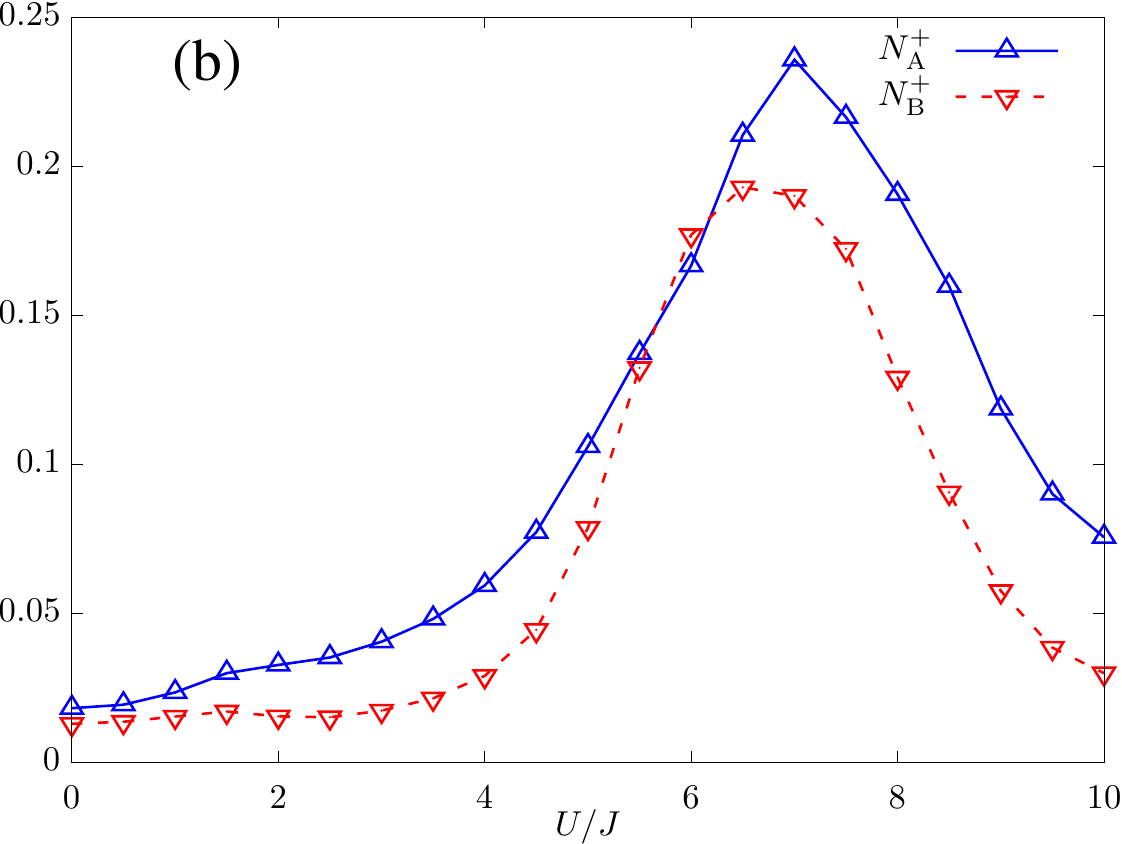}

\caption{\label{fig:cdw>0}Charge density with positive frequency induced by resonant tunneling for (a) $\Delta=-0.6J$ and (b) $\Delta=0$, plotted versus interactions $U$. The peak is due to one-photon resonant tunneling when $U=\Omega$. Other parameters
are the same as in Fig.~\ref{fig:cdw}. }
\end{figure}

There is a rich relation between charge density and interactions.
We show the charge density $N_{\mathrm A\left(\mathrm B\right)}$ (lines with empty up triangle and empty down triangle) with increasing
interactions in Fig.~\ref{fig:cdw}. (i) When $U=0$, the detuning
$\Delta$ is the factor that affects occupation of A and B sites,
and one can tune the occupations by changing $\Delta$.
(ii) We next discuss the case where $0<U\apprle3J$. The repulsive
interaction $U$ counteracts the effect of $\Delta$, and the density difference is reduced. In this region, the resonant tunneling is suppressed because the bandwidth is smaller than the driving frequency.
We show the spectral functions for $U=3J$ in Fig.~\ref{fig:Spectral}(b),
and we observe that the band width is approximately $\Omega$. The bandwidth
is smaller than $\Omega$ for a smaller $U.$ (iii) When $U\apprge3J$,
resonant tunneling plays an important role.  
We define $N_{\mathrm B\left(\mathrm A\right)}^{+}=\frac{1}{2\pi}\int_{-\Omega/2}^{\Omega/2}d\omega\sum_{n,\omega+n\Omega>0}\mathrm{Im}G_{nn,\mathrm B\left(\mathrm A\right)}^{<}\left(\omega\right)$, which corresponds to the fraction of atoms occupying the upper Mott band. The reason for these excitations is the resonant tunneling induced by the hopping $\left\langle F\right|e^{i\tilde{\theta}_{\mathrm B\mathrm A}\left(t\right)}c_{\mathrm B}^{\dagger}c_{\mathrm A}\left|I\right\rangle $  between A and B sites in the correlated regime. 
If there were no such resonant tunneling, the fraction of atoms on A site
would be estimated as $N_{\mathrm A}-N_{\mathrm A}^{+}+N_{\mathrm B}^{+}$, and similarly
for B site it would be $N_{\mathrm B}+N_{\mathrm A}^{+}-N_{\mathrm B}^{+}$. If the estimation
were accurate, we would have $N_{\mathrm A}-N_{\mathrm A}^{+}+N_{\mathrm B}^{+}\approx N_{\mathrm B}+N_{\mathrm A}^{+}-N_{\mathrm B}^{+}\approx 0.5$
for interactions $U\apprge3$. As shown in Fig.~\ref{fig:cdw}(a) and (b),
we note that both $N_{\mathrm A}-N_{\mathrm A}^{+}+N_{\mathrm B}^{+}$ (solid up triangle) and $N_{\mathrm B}+N_{\mathrm A}^{+}-N_{\mathrm B}^{+}$ (solid down triangle)
are close to 0.5 when $U\apprge3J$. This demonstrates that the resonant
tunneling between A and B sites is the main contribution to the difference between the atom densities on A and B sites when $U$ is relatively large.
This shows the characteristic difference between the NESS and an equilibrium state. On
the other hand, it offers a way to estimate the resonant tunneling.
Suppose it is possible to measure the charge density $N_{A}$ and
$N_{B}$ for different interactions. One can then estimate the contribution
of the resonant tunneling by calculating $N_{\mathrm A}-0.5$ or $0.5-N_{\mathrm B}$
for $U\apprge3J$. (iii) There is some deviation from 0.5 for
$N_{\mathrm A}-N_{\mathrm A}^{+}+N_{\mathrm B}^{+}$ and $N_{\mathrm B}+N_{\mathrm A}^{+}-N_{\mathrm B}^{+}$. It
shows that there are high-order contributions, from the hopping
A-B-A (or equivalently, B-A-B), to $N_{\mathrm A}^{+}$and $N_{\mathrm B}^{+}$. These
contributions do not affect the number of atoms on A and B sites,
but create excitations. This is also evidence for that the next-nearest
neighbor hopping is effectively generated by lattice shaking, even
though in this calculation we cannot show that it comes with a phase in this calculation. 

We have a short comment regarding the case in Fig.~\ref{fig:Spectral}(a).
When the staggered potential is absent, there are still resonant tunnelings.
However, they are the same for A and B sites, in contrast to the case with staggered potentials.

In Fig.~\ref{fig:cdw>0}, the charge density with positive frequency ($N_{\mathrm A\left(\mathrm B\right)}^{+}=\frac{1}{2\pi}\int_{-\Omega/2}^{\Omega/2}d\omega\sum_{n,\omega+n\Omega>0}\mathrm{Im} G_{nn,\mathrm A\left(\mathrm B\right)}^{<}\left(\omega\right)$)
is shown. It is corresponding to the effect of resonant tunneling~\cite{Qin:2017aa}.
The peak is at $U=\Omega=7J$, where the one-photon resonant tunneling
is dominant. As we have discussed above, the contribution $N_{\mathrm{A}}^{+}$
($N_{\mathrm{B}}^{\dagger}$) is due to the direct hopping from B (A) sites, plus higher-order
contributions from A (B) sites. Generally, $N_{\mathrm{A}}^{+}>N_{\mathrm{B}}^{+}$, and
atoms prefer to hop to A sites due to the higher staggered potential on B sites.

\subsection{Edge states}

\begin{figure}[h]
\includegraphics[scale=0.85]{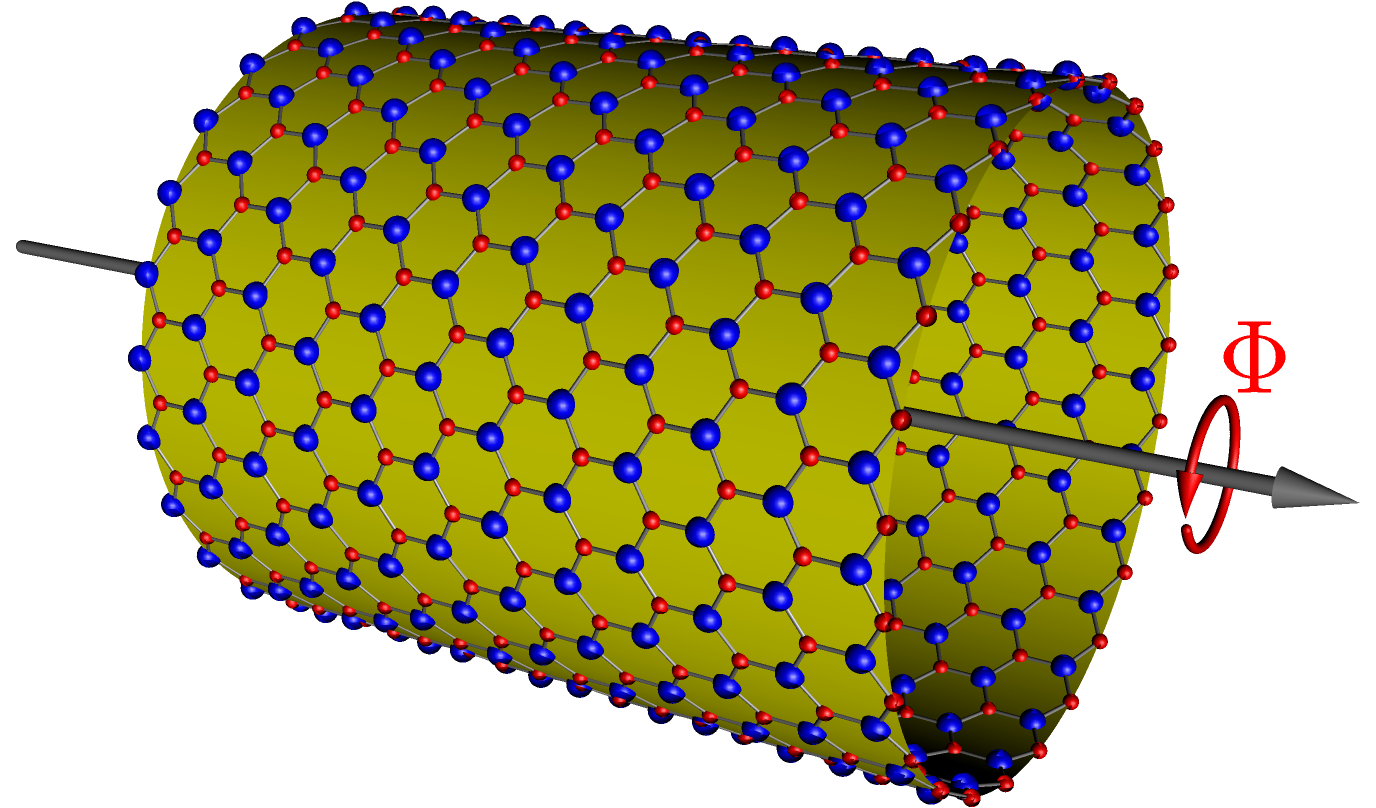}

\caption{\label{fig:cylinder}A cylinder geometry of a circularly shaken hexagonal
optical lattice consisting of A (big blue sphere) and B (small red sphere) atoms with a flux $\Phi$ insertion. This is the setup in
the Laughlin gedanken experiment~\cite{Laughlin1981}. The zigzag
boundary condition is used in the calculation. When there are well-defined edge states, they are expected
to be present in the left and the right edges.}

\end{figure}

In this section, we present topological properties of a circularly
shaken hexagonal optical lattice. We investigate edge states in a
cylinder geometry of a hexagonal optical lattice, with a flux $\Phi$
insertion (Fig.~\ref{fig:cylinder})~\cite{Wang2014tm,Grushin2015csf,Zaletel2014fie}.
This is the setup in the Laughlin gedanken experiment~\cite{Laughlin1981}.
The insertion of flux is equivalent to a twisted boundary condition
in the direction with periodic boundary condition~\cite{Niu1985qi,Qi2006gi}.
It is a general setup with the potential to be generalized to disordered
cases~\cite{Qi2006gi}. 

\begin{figure}[h]
\includegraphics[scale=0.99]{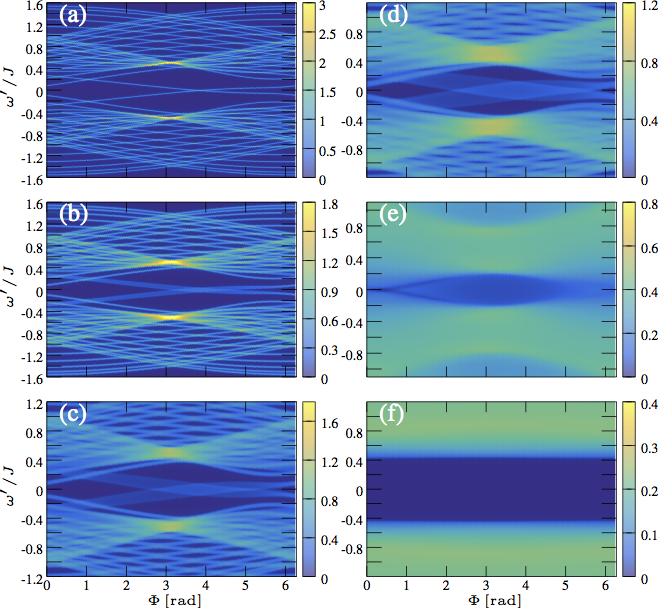}

\caption{\label{fig:edges}Spectral functions (with unit [TL$^{-2}$]) versus the flux $\Phi$ for different
interactions. Explicitly, (a) $U=0$, (b) $U=0.1J$, (c) $U=0.2J$, (d) $U=0.3J$, (e) $U=0.7J$,  and (f) $U=3J$. It shows interaction effects on edge states in a cylinder
geometry of $3\times10$ unit cells with $3$ unit cells in the direction
with a periodical boundary condition. Other parameters are $\Omega=7J$, $\frac{K}{\Omega}=1.28$, and
$\Delta=-0.35J$. Bath parameters are $\Gamma=0.005J$ and $T=0.01J$. $\omega^\prime$ indicates a general frequency which can be outside the first Brillouin zone of $\Omega$.}

\end{figure}

With insertion of one flux quantum, we demonstrate the change of topological
properties of the system. Edge states are hallmarks of nontrivial
topological properties. Using real-space Floquet DMFT, we study
the interplay between interactions and edge states. We show the spectral
functions in Fig.~\ref{fig:edges}. For finite detuning
$\Delta$, the cylinder geometry can host edge states (lines around $\omega^\prime=0$ in Fig.~\ref{fig:edges}(a)) when $U=0$. With increasing
interactions, the edge states as a function of the inserted flux $\Phi$ are smeared out as can be seen in Fig.~\ref{fig:edges}(b)-(d). This corresponds to
a finite lifetime of quasiparticles. The sharp spectral peak of a quasiparticle
is gradually expanded due to increasing interactions. When $U=3J$,
we see a simple Mott gap (Fig.~\ref{fig:edges}(f)). We observe three different phases: Chern
insulator with edge states present, pseudogap metallic phase with gap
closed, and Mott insulator with gap open again. According to the effective Hamiltonian~\cite{Tarnowski:2017aa}, the ratio between the next-nearest neighbor hopping $t_2$ and nearest neighbor hopping $t_1$ is $|t_2/t_1|=0.1$ with $|t_1|=0.5J$. The pseudogap metallic phase exists approximately when $U\approx 2J$. This is consistent with
DMFT calculations in Ref.~\cite{Nguyen2013cdp} for the Haldane-Falicov-Kimball
model. The difference is that here we are considering a non-equilibrium
driven system connected to a bath. The largest interaction $U=3J$
we have shown is much smaller than driving frequency $\Omega=7J$.
It can therefore be expected that resonant tunneling is greatly suppressed.
The dissipation rate $\Gamma$ into the bath has effects on the spectral
functions especially for small $U$. It introduces an $i\Gamma$ correction to the self-energy
and this term is equivalent to an interaction effect. 

\subsection{Charge pump }

\begin{figure}[h]
\includegraphics[scale=0.85]{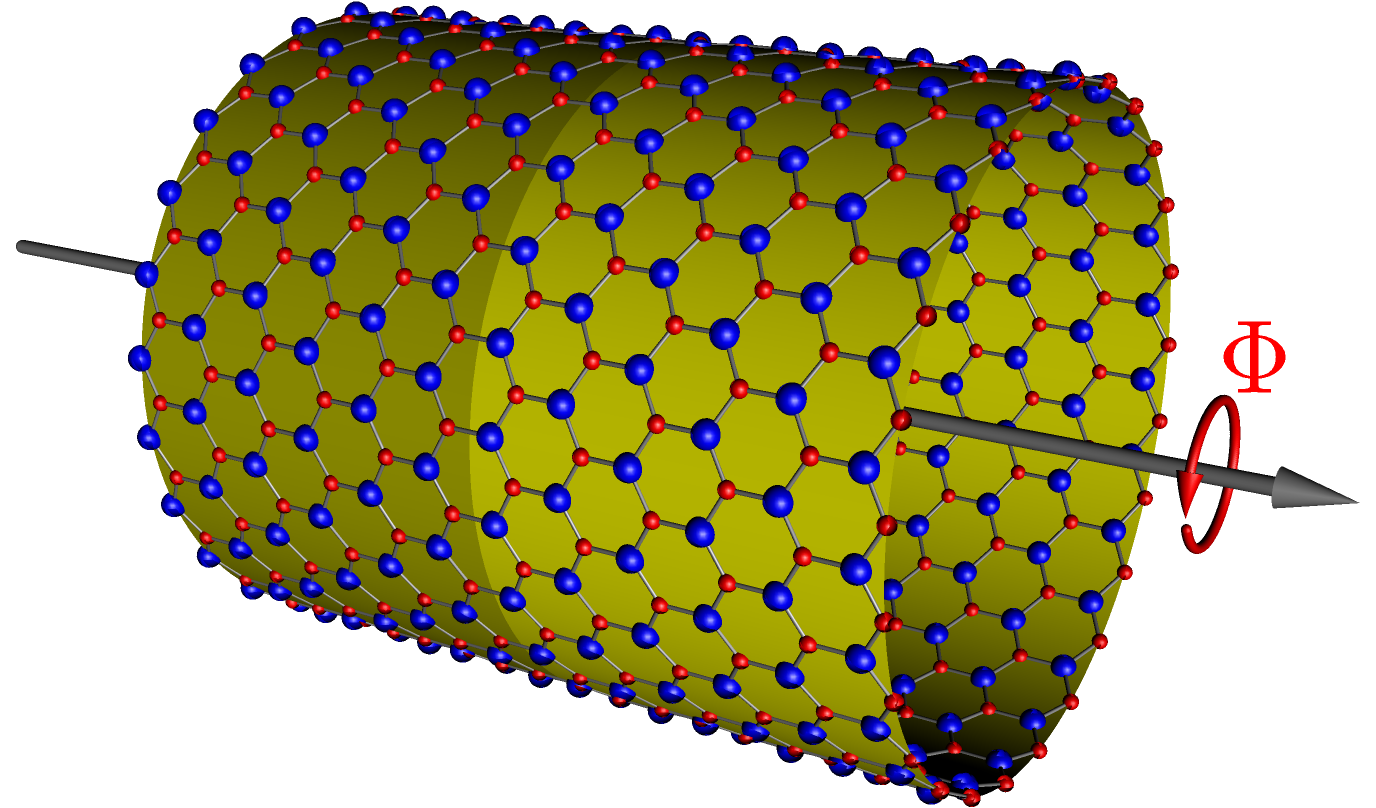}

\caption{\label{cylinderpump}The charge pump with insertion of a flux $\Phi$ is defined as charge
density difference $Q_{\Phi}=Q_{\mathcal{R},\Phi}-Q_{\mathcal{L},\Phi}$
between the two halves of the cylinder, as indicated by different background
colors (shading and unshading).}
\end{figure}

A second topological quantity which can be investigated in Laughlin's
setup is the charge pump, which is closely related to edge states.  For an equilibrium system, when
well-defined edge states are present in the cylinder geometry of a 
hexagonal optical lattice, a change of one flux quantum will induce an integer
number of atoms to transfer from one edge of the cylinder to the other~\cite{Laughlin1981}.
The number of transferred atoms depends on the number of edge states.
An integer charge pump is a signature of non-trivial topological
phase. Following Ref.~\cite{Wang2014tm}, we define the charge pump with
insertion of flux $\Phi$ as
\begin{equation}
Q_{\Phi}=Q_{\mathcal{R},\Phi}-Q_{\mathcal{L},\Phi}\label{eq:chargepump}
\end{equation}
which is the charge density difference between two halves of the cylinder
(see Fig.~\ref{cylinderpump}: left (shading) and right (unshading) halves of the cylinder). $Q_{\alpha,\Phi}=\frac{1}{2\pi}\int_{-\Omega/2}^{\Omega/2}d\omega\sum_{i\in\alpha,n}\mathrm{Im}G_{i,\Phi,nn}^{<}\left(\omega\right)$
with $\alpha=\mathcal{R},\mathcal{L}$, and Floquet index $n$.
$\mathrm{Im}G_{i,\Phi,nn}^{<}\left(\omega\right)=\frac{1}{2}\mathrm{Im}\left[G_{i,\Phi,nn}^{K}\left(\omega\right)-G_{i,\Phi,nn}^{R}\left(\omega\right)+G_{i,\Phi,nn}^{A}\left(\omega\right)\right]$,
where $i$ is the site index in the left or right half. The flux is
implemented according to Ref.~\cite{Zaletel2014fie}. Using the sum
rule $-\frac{1}{\pi}\int_{-\Omega/2}^{\Omega/2} d\omega\mathrm{Im}\sum_{n}G_{i,\Phi,nn}^{R(A)}=1(-1)$,
we have 
\begin{equation}
Q_{\Phi}=\int_{-\Omega/2}^{\Omega/2}d\omega\sum_{n}\frac{1}{2\pi}\mathrm{Im}\left[G_{\mathcal{R},\Phi,nn}^{K}\left(\omega\right)-G_{\mathcal{L},\Phi,nn}^{K}\left(\omega\right)\right].
\end{equation}

\subsubsection{An isolated equilibrium system}
\begin{figure}[h]
\includegraphics[scale=0.08]{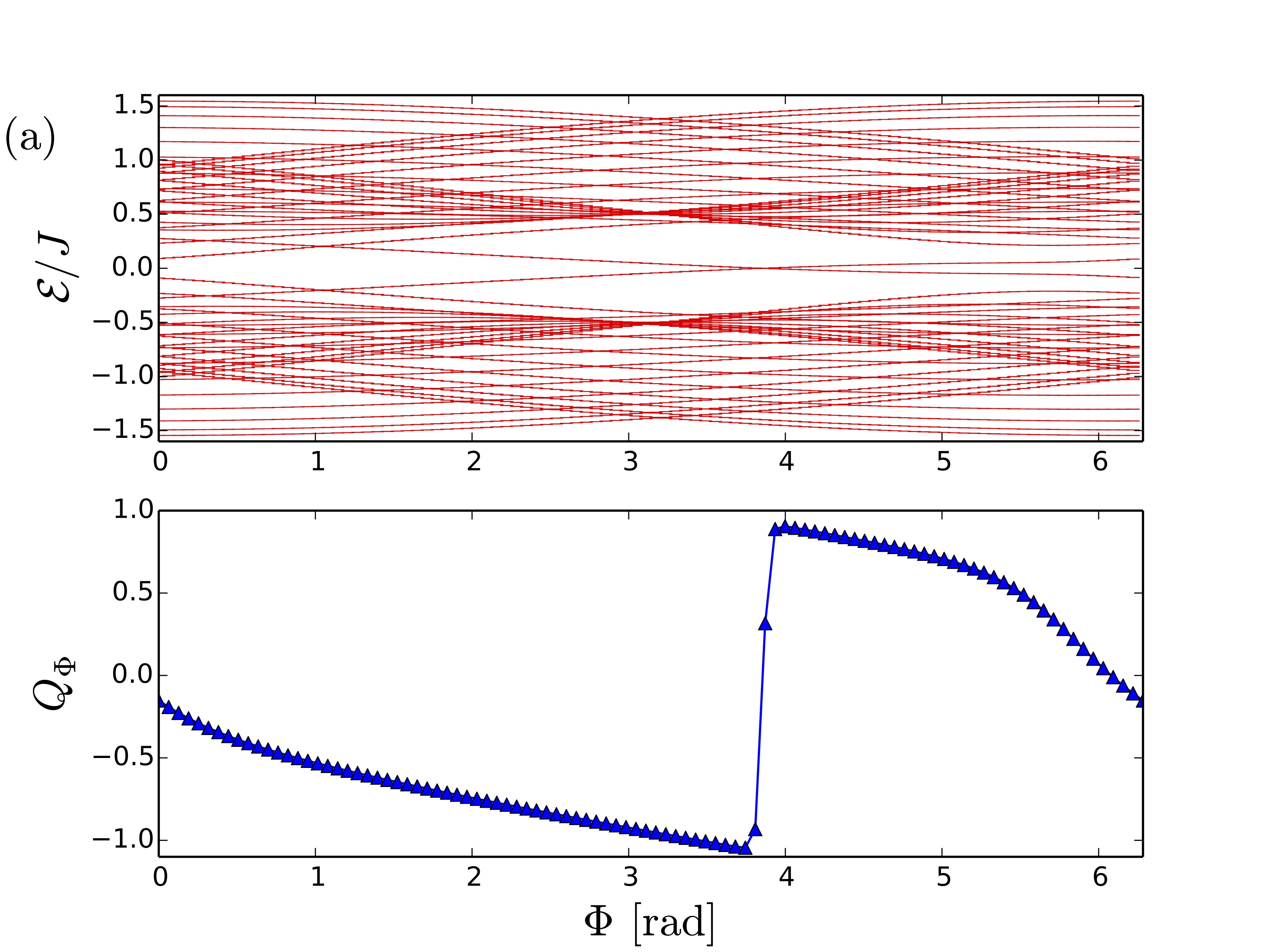}

\includegraphics[scale=0.08]{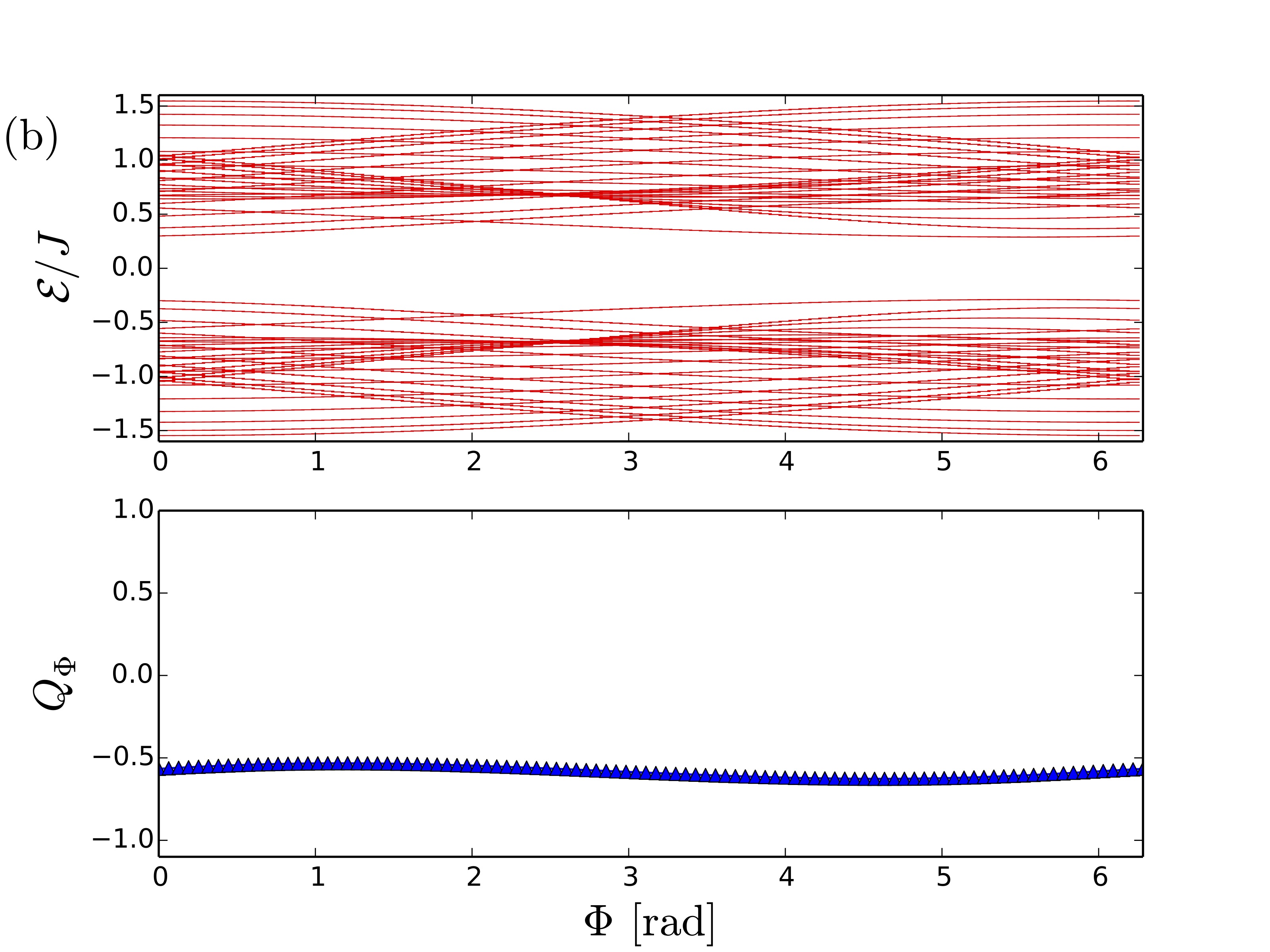}

\caption{\label{fig:charge_eq} Energy spectrum $\mathcal{E}$ and charge pump $Q_{\Phi}$ with insertion of a
flux for a cylinder hexagonal lattice with $3\times10$ unit cells
in \emph{an non-interacting equilibrium state}. There are 3 unit cells
in the direction with periodic boundary conditions. Other parameters  are $\Omega=7J$,
$\frac{K}{\Omega}=1.28$, $\Delta=-0.35J$ (panel (a)), and $\Delta=0.35J$
(panel (b)).}

\end{figure}

We show that $Q$ can indeed be a topological invariant to
distinguish non-trivial and trivial topological phases for \emph{an
non-interacting equilibrium state}. For this case, no bath is needed
for energy dissipation. We can determine the Keldysh Green's function
using the fluctuation-dissipation theorem~\cite{rammer2007quantum}:
$G^{K}\left(\omega\right)=\tanh\frac{\beta_{0}\omega}{2}\left(G^{R}\left(\omega\right)-G^{A}\left(\omega\right)\right)$,
where $\beta_{0}=\frac{1}{k_{B}T_{0}}$ with $T_{0}$ the equilibrium
temperature of the system. We choose $T_0$ as a very small number close to 0. The charge pump can be calculated as:
\begin{align}
Q_{\Phi}= & -\frac{1}{\pi}\mathrm{Im}\int_{-\Omega/2}^{\Omega/2}d\omega\sum_{n}f\left(\omega+n\Omega\right)\nonumber \\
 & \times\mathrm{Im}\left[G_{\mathcal{R},\Phi,nn}^{R}\left(\omega\right)-G_{\mathcal{L},\Phi,nn}^{R}\left(\omega\right)\right].
\end{align}
$Q_{\Phi}$ can be calculated directly or using the technique of the contour
integral. In Fig.~\ref{fig:charge_eq}, we show the charge pump versus
flux insertion $\Phi$ for topologically non-trivial and trivial
cases. In Fig.~\ref{fig:charge_eq}(a), we observe
a sharp jump of the charge density difference $Q$ at the point $\Phi$,
where two edge states intersect each other. It means that one atom
is transferred from one edge of the cylinder to the other. In
contrast, in the topologically trivial case (Fig.~\ref{fig:charge_eq}(b)), we only observe a smooth
change in the charge density. Therefore, $Q_{\Phi}$ can serve as
a topological invariant for distinguishing topologically non-trivial and
trivial cases. 

\subsubsection{System coupled to a free fermion bath}

\begin{figure}[h]
\includegraphics[scale=0.7]{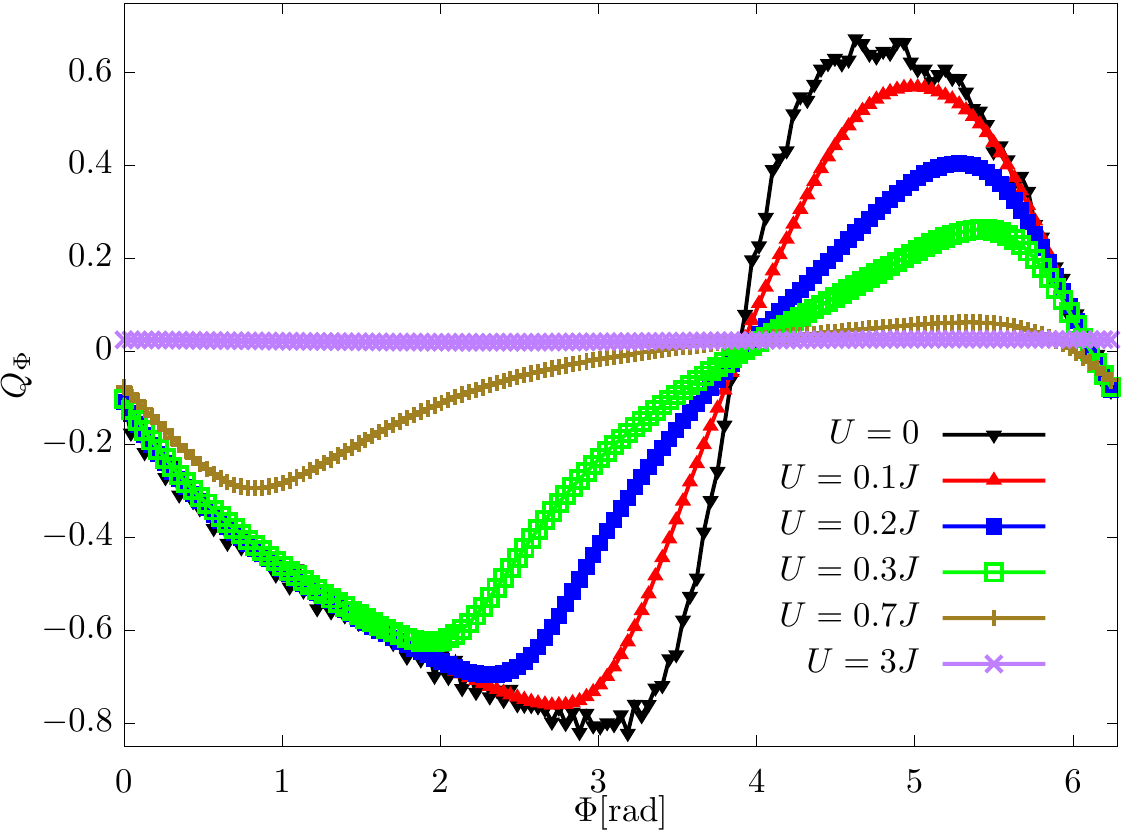}

\caption{\label{fig:charge_ness}Charge pump $Q_{\Phi}$ with insertion of a flux
$\Phi$ for a cylinder hexagonal lattice, obtained by real-space
Floquet DMFT for different interactions. The parameters are the same
as those in Fig.~\ref{fig:edges}. For the non-interacting case, we obtain the charge pump by a direct calculation of the non-interacting Floquet Green's function with the bath corrections. It is different from the method of rate equations in Sec.~\ref{lindbladma}.}

\end{figure}

We present the charge pump $Q$ for a NESS obtained from the real-space
Floquet DMFT in Fig.~\ref{fig:charge_ness}. Corresponding to edge
states in Fig.~\ref{fig:edges}, we show that there is a jump in
the charge pump in Fig.~\ref{fig:charge_ness}. However, the jump
is not an integer even for very small interactions. With increasing
interactions, the jump becomes very smooth. We explain why the jump
is not integer even for $U=0$ (line with solid down triangle). It does not contradict to what we show in
Fig.~\ref{fig:charge_eq}. In real-space Floquet DMFT, there
is a bath coupled to every lattice site. With the approximation of constant DOS of the bath, the dissipation into the bath
introduces a finite self-energy to the system. This is equivalent to
an interaction effect. In fact, it is this effective interaction which destroys the integer
charge pump when $U=0$. When the system couples to the environment (the bath) and becomes open, the unavoidable dissipation plays a role in the topological properties. Even though the dissipation is rather small,  the interaction effect induced by it can be pronounced because the effective hopping is heavily dressed by the driving. 

\begin{figure}[h]
\includegraphics[scale=0.7]{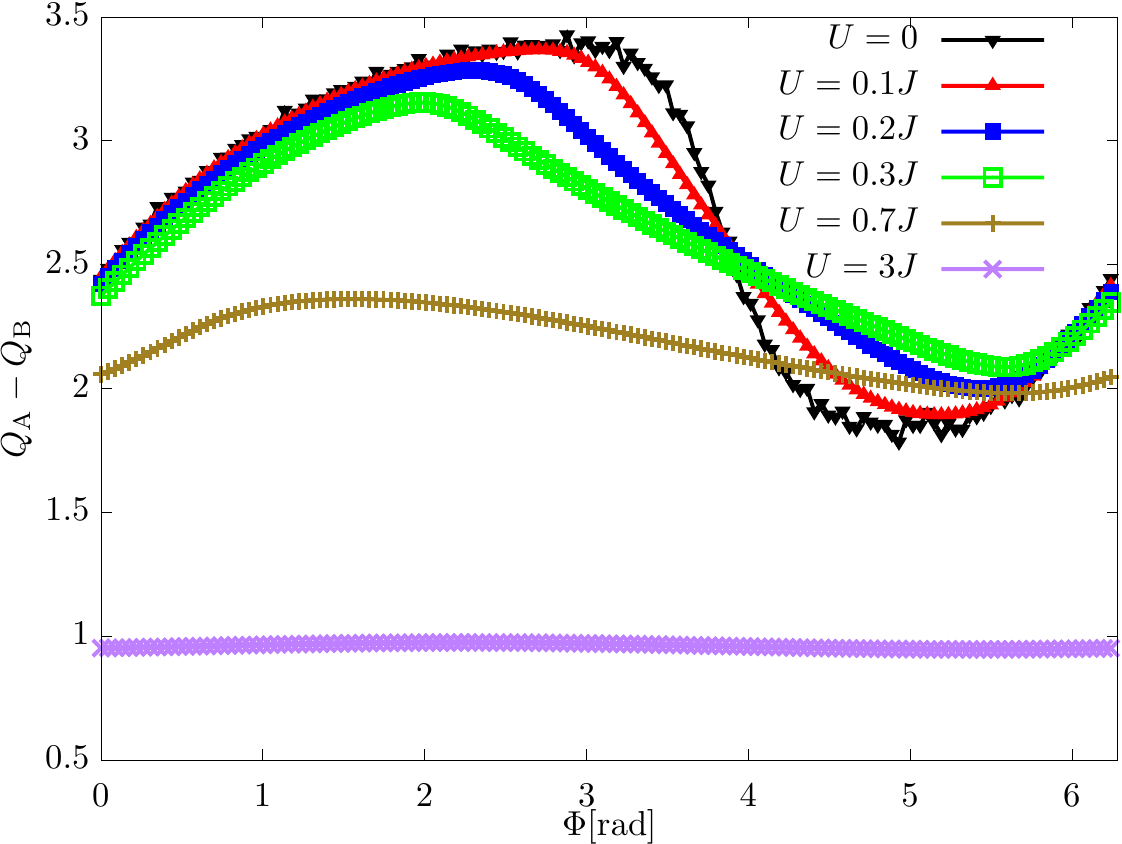}

\caption{\label{fig:cdw_phi}The total charge density difference $Q_{A}-Q_{B}$ on a cylinder hexagonal lattice for the (interacting)
Floquet case coupled to a bath. The parameters are the same
as those in Fig.~\ref{fig:edges}. A jump is observed for $U=0$ (line with solid down triangle).}

\end{figure}

A connection can be made between the charge pump and CDW in Sec.~\ref{sec_cdw}~\cite{Grusdt2014rb}. We observe a jump in both Figs.~\ref{fig:charge_ness} and~\ref{fig:cdw_phi}  for $U=0$, which signals a possible relationship.
For the cylinder geometry threaded by flux, we can describe the CDW
$Q_{\mathrm{A}}-Q_{\mathrm{B}}=Q_{\mathcal{R},\mathrm{A}}+Q_{\mathcal{L},\mathrm{A}}-Q_{\mathcal{R},\mathrm{B}}-Q_{\mathcal{L},\mathrm{B}}$,
which is the total density difference between site A and B. The charge
pump is defined by $Q_{\mathcal{R}}-Q_{\mathcal{L}}=Q_{\mathcal{R},\mathrm{A}}+Q_{\mathcal{R},\mathrm{B}}-Q_{\mathcal{L},\mathrm{A}}-Q_{\mathcal{L},\mathrm{B}}$.
A simple comparison shows that the difference
lies in the part $Q_{\mathcal{L},\mathrm{A}}-Q_{\mathcal{R},\mathrm{B}}$.  In Fig.~\ref{fig:cdw_phi} we note that charge density difference between A and B exists at $\Phi=0$ due to CDW, and we see a clear jump when $\Phi\approx3.8\mathrm{[rad]}$ for $U=0$ (line with solid down triangle), exactly where the jump happens for the charge pump in Fig.~\ref{fig:charge_ness}. For the equilibrium case, see Ref.~\cite{Supple2018}. Therefore, for a pronounced charge pump, there must be a dominated hopping process from A to B or from B to A, which can lead to a significant density redistribution between A and B.  Furthermore, our discussion in Sec.~\ref{sec_cdw}
shows that the weak interaction counteracts the effects of staggered
potentials, and makes the density difference between A and B smaller.
Consistently, in Fig.~\ref{fig:cdw_phi}  we see that changes in charge density difference versus $\Phi$ becomes smaller with increasing interactions.

\subsubsection{System coupled to a heat bath}\label{lindbladma}
\begin{figure}[t]
\captionsetup[subfloat]{captionskip=-0.6cm, position=top, font=normalsize, justification=raggedright, singlelinecheck=off}
\subfloat[]{\hspace{0.5cm}\includegraphics[scale=0.8]{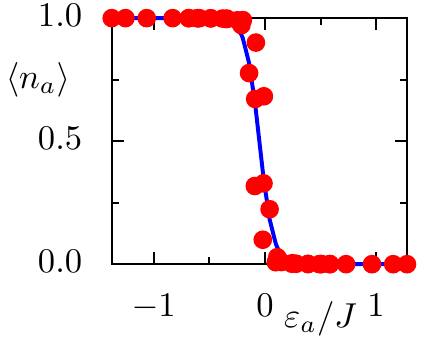}}
\subfloat[]{\hspace{0.5cm}\includegraphics[scale=0.8]{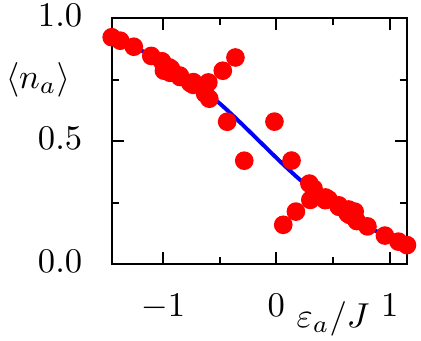}}\\ 
\subfloat[]{\hspace{0.5cm}
\includegraphics[scale=0.8]{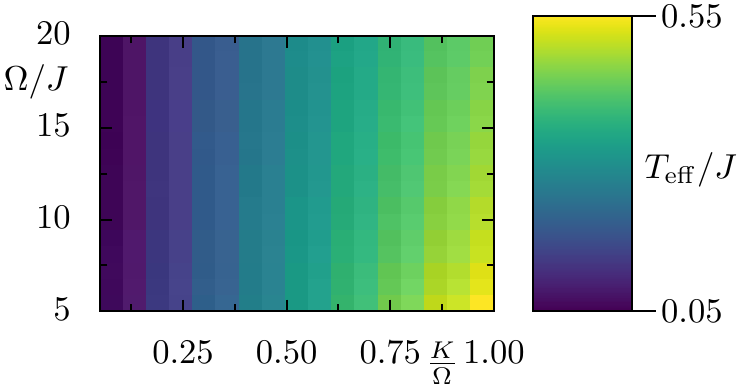}} 
\captionsetup[subfloat]{captionskip=-0.6cm, position=top, font=normalsize, justification=raggedright, singlelinecheck=off}
\subfloat[]{\hspace{0.5cm}
\includegraphics[scale=0.8]{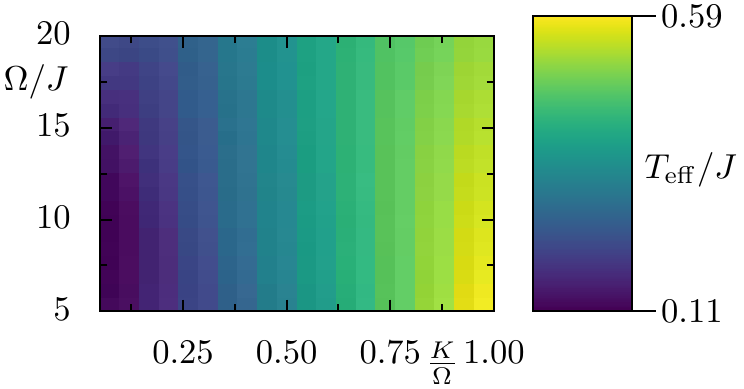}} 
\caption{\label{fig:heating_ness} NESS that forms when the system is  coupled to an ohmic 
heat bath (no cutoff, $E_c=\infty$) at $T=0.001J$. (a, b) We obtain effective temperatures by
fitting the distribution $\langle n_a\rangle$ (red dots) of the NESS with a Fermi-Dirac
distribution with temperature $T_{\mathrm{eff}}$ (blue solid line), where
(a) $\Delta=-0.1J$, $\Omega=5J$, $K/\Omega = 0.05$ and (b) $\Delta=-0.3J$, $\Omega=13.33J$, $K/\Omega = 1$. 
(c, d) $T_{\mathrm{eff}}$ as a function of driving parameters, for (c) 
 $\Delta=-0.1J$ and (d) $\Delta=-0.3J$. We show states for a $3 \times 7$ lattice, but
we observe that $T_{\mathrm{eff}}$ is almost independent of the size of the lattice.}

\end{figure}

To further identify the role of dissipation to the bath in the charge pump for a non-interacting system, we here study the NESS that
forms when the driven system at half filling is coupled to an ohmic heat bath, using the rate equations.

Similar to the fermionic reservoir, we couple one heat bath at temperature $T$
to every site of the lattice. This is mediated by a coupling operator  $v_{(l)} = c^\dagger_l c_l$
for a given site $l$. Note that we assume this form of the coupling for the direct frame. 
However, if we transform the coupling to the co-moving frame, it still obeys the same form 
since the unitary rotation $\mathcal{U}$ commutes with the coupling $c^\dagger_l c_l$.
From Eq.~\refeq{eq:rate-sp} one then infers rates $R^{(l)}_{ab}$ that
result from coupling this site $l$ to the heat bath. The total rates for coupling the system globally
to an external heat bath result from the incoherent sum of all of these processes, implying  $R_{ab} = \sum_l R^{(l)}_{ab}$.

With these rates we solve the kinetic equation \refeq{eq:kinetic} for the NESS. 
Just to remind the reader, the resulting state is the long-time steady state that results when the system is 
under a constant driving and weakly coupled to the bath, meaning that the coupling constant 
is small when compared to all quasi-energy splittings in the system, $\sqrt{\Gamma} \ll (\varepsilon_k - \varepsilon_q)$, for $k\neq q$.

We observe that for frequencies $\Omega$ which are large when compared to the bandwidth, the
distributions $\langle n_a \rangle$ that we observe in the NESS are still close to thermal distributions with an effective temperature $T_{\mathrm{eff}}$,
cf.~the examples in Fig.~\ref{fig:heating_ness}(a) and (b). This effective temperature is obtained by fitting the closest 
thermal distribution to the occupations such that $\langle n_a \rangle \approx 1/(\mathrm{e}^{(\varepsilon_a-\mu)/T_\mathrm{eff}}+1)$, and therefore assuming what 
was called a ``Floquet-Gibbs'' state in the literature~\cite{ShiraiEtAl16}.
Note that in Fig.~\ref{fig:heating_ness}, the
temperature of the bath is $T=0.001J$, but still, due to the driving, in the long-time limit the system heats up to quite
high temperatures that are on the order of $T_{\mathrm{eff}} \approx 0.1J$ as shown in Fig.~\ref{fig:heating_ness}(c) and (d)
for a heat bath with ohmic spectral density and no spectral cutoff, $J(E) \propto E$.
Interestingly, in this frequency regime, the effective temperature of the steady state seems to depend only on the
relative strength $K/\Omega$ of the driving. 
Note that this is in contrast to the analytic formula that was presented in Ref.~\cite{Iadecola2015b}, where
in addition to the $K/\Omega$ dependence they find a term that scales as $1/\Omega^d$ (where $d$ is the 
exponent of the spectral density).
In our calculations we also 
observe that $T_{\mathrm{eff}}$ is practically independent of the size of the lattice.

\begin{figure}[h]
\captionsetup[subfloat]{captionskip=-0.6cm, position=top, font=normalsize, justification=raggedright, singlelinecheck=off}
\subfloat[]{\hspace{0.5cm}\includegraphics[scale=0.8]{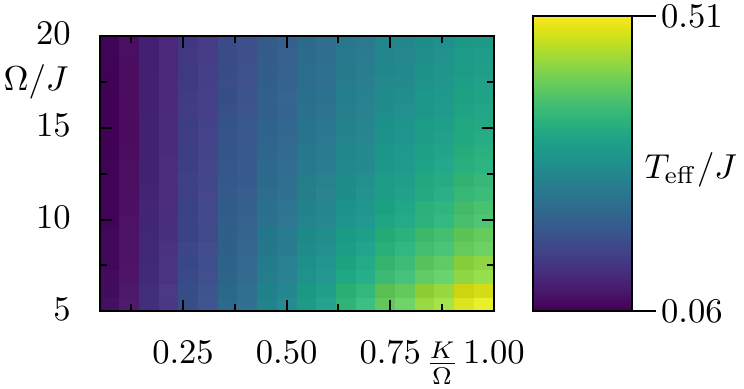}} \\
\subfloat[]{\hspace{0.5cm}\includegraphics[scale=0.8]{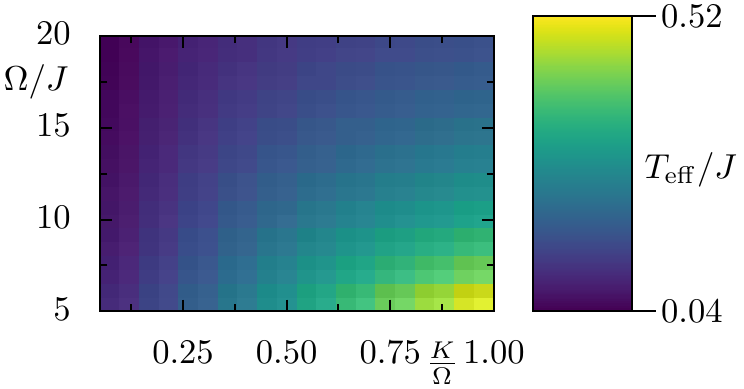}} 
\caption{\label{fig:heating_nonohmic} As in Figure \ref{fig:heating_ness}(d), but (a) with $J(E) \propto E^{0.5}$ and
(b) with $J(E) \propto E \exp(-E/5J)$.}
\end{figure}

\begin{figure}[h]
\captionsetup[subfloat]{captionskip=-0.6cm, position=top, font=normalsize, justification=raggedright, singlelinecheck=off}
\subfloat[]{\hspace{0.5cm}\includegraphics[scale=1]{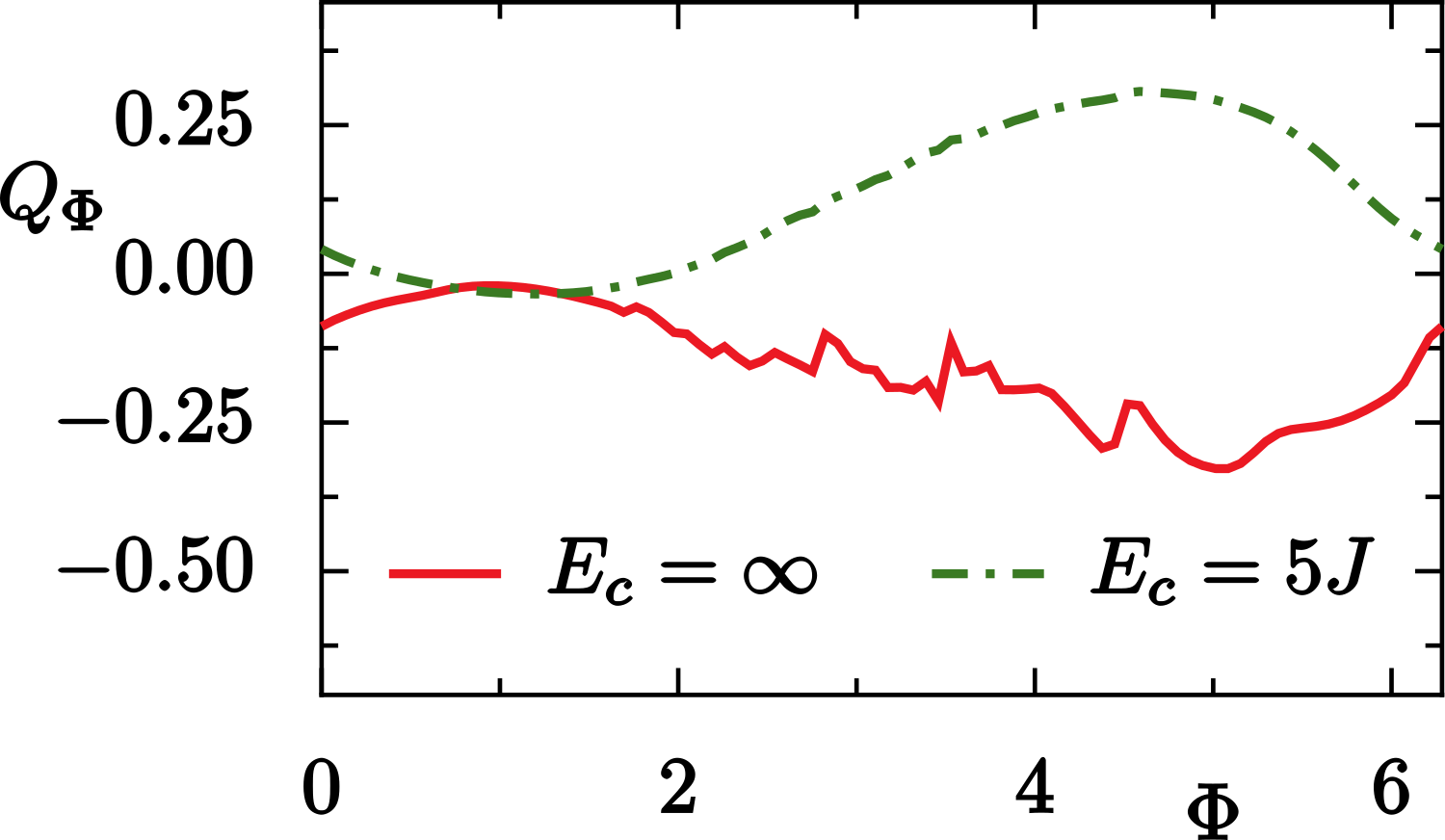}}\\
\subfloat[]{\hspace{0.5cm}\includegraphics[scale=1]{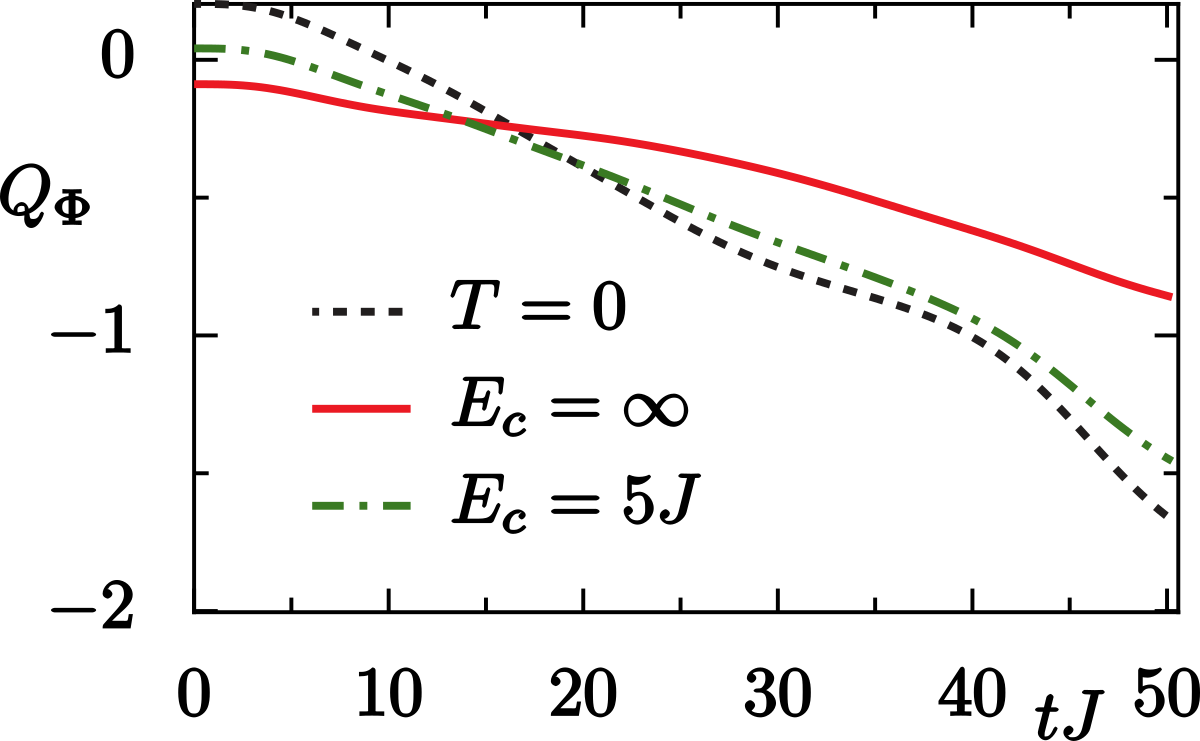}}
\caption{\label{fig:heating_chargepump} (a) Charge Pump $Q_\Phi$ with insertion of a flux $\Phi$
for parameters $\Omega=20J, \Delta=-0.1J, K/\Omega = 0.25$ in a $3\times10$ lattice where we expect the system to be topological.
For given flux $\Phi$ 
we let the system relax to the NESS in presence of a heat bath with $J(E) \propto E \exp(-E/E_c)$
and $T=0.001J$. For the green dash-dotted curve there is a spectral cutoff $E_c=5J$, the red solid curve is for ohmic bath.  Using this procedure the signal is very weak, 
even though the effective temperature in the NESS are quite low.
(b) However, there is a large transport signal if we prepare the NESS at $\Phi=0$ and then ramp (at constant speed) one flux quantum
within a finite ramp time $\tau\approx50/J$. 
Here, the red solid and green dash-dotted lines show snapshots of the system at integer multiples of $\mathcal{T}$ neglecting the micromotion.
The green dash-dotted line (with some cutoff in the spectral density) features a relatively large charge transport that is
close to the (almost)  integer transport behavior that we expect for hypothetic thermal $T=0$  populations of the quasienergies (black dashed line).}
\end{figure}

Even by further decreasing the temperature $T$ of the bath, we are not able to reach
lower effective temperatures $T_{\mathrm{eff}}$.
We checked this by comparing to the NESS for a hypothetic $T=0$ bath, 
where there are no bath occupations $n_B=0$, so that there is only spontaneous emission. 
These relatively large effective temperatures are detrimental for the observation of quantized charge pumping, since they correspond to a significant occupation of the ``upper'' Floquet band. An extremely low temperature is vital for the
exactness of the integer quantum Hall effect~\cite{Giuliani:2005aa}.
A finite temperature excites the particles across the gap, while a
low temperature can make this probability exponentially low.

Note that this heating is due to the population transfer between Floquet states that
is induced due to the presence of the coupling to the higher  Floquet sidebands.
As was pointed out in the literature \cite{Seetharam2015cb,Iadecola2015b, Dehghani2015oi}, this heating can be suppressed by engineering the bath such that the spectral density $J(E)$ at large
quasi-energy differences becomes smaller. For example if we suppose
the bath is sub-ohmic with $J(E) \propto E^{0.5}$, as shown in Fig.~\ref{fig:heating_nonohmic}(a), then at large
frequencies $\Omega$ we find that heating is suppressed, leading to lower effective temperatures $T_\mathrm{eff}$
in the NESS. Similar suppression of heating is found in Fig.~\ref{fig:heating_nonohmic}(b) for an ohmic bath,
but with a finite cutoff $E_c = 5J$ in the spectral density $J(E) \propto E \exp(-E/E_c)$. 
Note that it  has been argued in the literature that in the limit $\Omega \gg E_c$ resonances are suppressed
and one expects an effectively thermalized ``Floquet-Gibbs'' state~\cite{ShiraiEtAl16}.
Such a finite frequency cutoff
is the manifestation of  bath correlation times $\tau_R \propto 1/E_c$ that are on the order of the time scales of the system dynamics
 $\tau_S \propto 1/J$. 
 Note that such finite correlation times are tunable e.g.~in the case where the bath is a weakly interacting Bose-Einstein condensate
 in a trap.
There, excitations are nicely described by Bogoliubov quasiparticles (phonons) \cite{OzeriEtAl2005}
and the bath correlation times can be controlled by the trap frequency \cite{KleinFleischhauer2005, ostmann2017cooling}. 
Also, sympathetic cooling of fermions in Bose-Einstein condensates is a well established experimental technique, however
there,  typically a relatively strong coupling $\Gamma$  is favorable, while here we target weak couplings.

If we now use a  NESS that was prepared in presence of  such a heat bath, one again may ask whether one can observe
the underlying topological nature of the model. 
Similar to the DMFT calculations in presence of the fermionic reservoir, in Fig.~\ref{fig:heating_chargepump}(a) we show
the charge difference $Q_\Phi$ of the right and lefthand side of the system in the NESS that was prepared for a given value of the flux $\Phi$.
However, even though for these parameters the model is topological and the effective temperatures are relatively low,
especially in the case with a finite cutoff in the spectral density (green dash-dotted line), we do not see a pronounced peak
in the charge difference.  

In order to overcome this problem,
 we propose a different strategy to probe topology in the model. Namely, in Fig.~\ref{fig:heating_chargepump}(a)
it is assumed that the ramp time $\tau$ is big when compared to the relaxation times $\tau_R$, i.e.~we follow the system adiabatically
in the thermodynamic sense.
Here, since we are in the weak coupling regime where $\tau_R$ is large, we propose to perform a ramp  on a much shorter time scale
 $\tau \ll \tau_R$, such that during the ramp one may neglect the action of the bath. However,  this ramp should still be slow
 when compared to system time scales $\tau_s \ll \tau$, which one might call adiabatic in the closed system (without the presence of the bath).
The red solid and green dash-dotted lines in Fig.~\ref{fig:heating_chargepump}(b) show the charge transport that one observes in such a procedure, 
where we start with a NESS that is prepared without the presence of an external field, $\Phi_{\mathrm{NESS}} = 0$. There, 
the values are quite high, for the ohmic bath with a cutoff. The density difference between two halves of the cylinder  $Q_{\Phi}$ is up to about 1.4 (green dash-dotted line), corresponding to $0.7$ of a charge is transported, which is very close
to the quantized value that we expect for such a ramp if we  assume a $T=0$ population of the quasi-energies (black dashed line).

\subsection{Experimental relevance }

We discuss the possibilities to observe the physical quantities we
have explored in an experiment. 
To detect the CDW, one can measure the local densities on A and B sites either in situ in a quantum gas microscope, after time of flight via adiabatic band mapping techniques, or via the double occupancy~\cite{Messer2015ef}.
For the charge pump, one needs
to compare the local particle density between two parts of the cylinder geometry.
The extra flux, in fact, relaxes the requirement to have a periodic
boundary condition in one direction. For a hexagonal lattice which
is finite in both directions, if it is possible to connect the sites
at the ends of one direction with a complex long-range hopping, we
can realize this cylinder geometry with a flux. This might be easier using a synthetic dimension~\cite{Boada2015qs}. A second possible
way is discussed in Ref.~\cite{Lkacki2016qhp}. It proposes to use
Laguerre-Gauss beams to create a cylinder optical lattice. Another possibility is to engineer a ring shaped system with the central hole pierced by a tunable magnetic flux, as it can be realized using the scheme proposed in Ref.~\cite{Wang:2018aa}. For the charge pump measurement, as we have shown for the non-interacting case with rate equations, there is the possibility to 
couple the system  weakly to a low-temperature heat bath to prepare the system in a state close to equilibrium.
Bath engineering can be used to reach sufficiently low effective temperatures
in the NESS.
 Then, the flux can be adiabatically ramped up to have a pronounced charge pump. 
 
 The Falicov-Kimball
interaction can be realized in different possible ways~\cite{Eckstein2009,Jotzu2015a}. By introducing two species of atoms to
the optical lattice, one species of atoms can be localized by a deep
optical lattice depth~\cite{Eckstein2009}. The localized atoms are
in annealed disorder state for the Falicov-Kimball model, in contrast
to the quenched disorder realized in ultracold atoms. It may be possible
to be realized by switching off the hopping of the localized atoms
slowly. A second possibility is presented in Ref.~\cite{Jotzu2015a}
where the hopping of one species can be turnt off by tuning the driving
amplitude because different species experience different driving amplitudes. 
 
 We have a comment on the bath. To study NESS in a driven system, it is necessary to connect the system to a bath to dissipate the extra energy. Most setups for cold atoms in optical lattices are isolated systems, but they can be in the prethermal regime when the driving frequency is sufficiently large. We may expect that the NESS in our setup may share some similarities for an isolated system in the prethermal regime~\cite{Else2017ppm,Kuwahara2016fmt,Mori2016rbe}. 

\section{\label{sec:Cc}Conclusion}

In conclusion, we have studied the charge density wave and charge pump
of fermions with Falicov-Kimball interactions in a circularly shaken hexagonal
optical lattice. We show that the charge density wave is induced by the staggered potential and is dramatically
changed because of resonant tunneling. We also show interaction effects on the topological properties. An increase of the Falicov-Kimball interaction tends to smear out the edge states, and finally makes the system enter the Mott insulator phase. 

Furthermore, we study non-equilibrium steady states
 in a Laughlin charge pump setup  which is coupled to either a fermionic reservoir or a heat bath.
 In the interacting case, we show
that the charge pump is not integer for insertion of one flux quantum. 
Also for the non-interacting
case, we find that it is not integer due to dissipation into a bath. We confirm this by detailed calculations via rate equations based on the Floquet-Born-Markov approximation. Moreover, we explored possibilities to lower the effective temperature characterizing the NESS of the driven system by engineering the spectral properties of the bath.
Our calculations suggest that in theory one  can indeed use the presence of a bath to cool down the system,
e.g. after a quench where in the closed system typically there are excitations in the upper band, and also to some
extent one can overcome the heating that is inherent in the interacting Floquet system. 
We propose an experimentally feasible procedure to ramp up the flux for the measurement of the charge pump.  

In the future, the approaches developed here can also be applied to the Haldane-Hubbard model, where both spin states are mobile, and which is naturally realized with cold atoms in optical lattices.

\section{Acknowledgments}
This work is supported by the Deutsche Forschungsgemeinschaft via
DFG FOR 2414 and the high-performance computing center LOEWE-CSC.
The authors acknowledge useful discussions and communication
with M. Eckstein, K. Le Hur, and N. Tsuji.
\bibliographystyle{apsrev4-1}
%


\newpage

\onecolumngrid
\setcounter{figure}{0} \setcounter{equation}{0} \renewcommand\theequation{S\arabic{equation}}

\noindent
\begin{center}
\large{\bf Supplementary material}
\end{center}


\subsection{Impurity solver for the Falicov-Kimball model}

We outline how to include the Falicov-Kimball interaction in the impurity
part following Refs.~\cite{sEckstein2008nss} and~\cite[chapter 5.2]{sEckstein2009}
in the formalism of nonequilibrium DMFT. 
\begin{enumerate}
\item In DMFT, for a given site, local correlation functions are obtained
from an effective single-site problem with action
\begin{align}
\mathcal{S}= & \mathcal{S}_{\mathrm{loc}}+\mathcal{S}_{\mathrm{hyb}}\\
\mathcal{S}_{\mathrm{loc}}= & -i\int_{\mathcal{C}}dtH_{\mathrm{loc}}\left(t\right)-i\mu\sum_{\sigma}\int_{\mathcal{C}}dtc_{\sigma}^{\dagger}\left(t\right)c_{\sigma}\left(t\right)\\
\mathcal{S}_{\mathrm{hyb}}= & -i\sum_{\sigma\sigma^{\prime}}\int_{\mathcal{C}}dt\int_{\mathcal{C}}dt^{\prime}c_{\sigma}^{\dagger}\left(t\right)\Lambda_{\sigma\sigma^{\prime}}\left(t,t^{\prime}\right)c_{\sigma^{\prime}}\left(t^{\prime}\right)
\end{align}
where $\mathcal{S}_{\mathrm{loc}}$ contains the dynamics induced
due to the local Hamiltonian $H_{\mathrm{loc}}\left(t\right)=U\left(t\right)n_{\uparrow}\left(t\right)n_{\downarrow}\left(t\right)$.
For the Falicov-Kimball interaction, one can imagine one spin species
as the $f$-particle. 
\item For the Falicov-Kimball model, in DMFT there is an important simplification
for the action $\mathcal{S}$ due to the immobility of the $f$-particles:
$\Lambda_{ff}\left(t,t^{\prime}\right)=0$. In fact, $\Lambda_{ff}\left(t,t^{\prime}\right)$
describes temporal fluctuations of the $f$-particle density. We can
replace the density operator $f^{\dagger}\left(t\right)f\left(t\right)$
by the time independent operator $n_{f}$ in the action, and we have
the local Green's function for a general time-dependent interaction
$U\left(t\right)$:
\begin{align}
G\left(t,t^{\prime}\right)= & -i\frac{\mathrm{Tr}_{c,f}\left[e^{-\beta H_{0}}\mathrm{T}_{\mathcal{C}}\exp\left(\mathcal{S}\left[n_{f}\right]\right)c\left(t\right)c^{\dagger}\left(t^{\prime}\right)\right]}{\mathrm{Tr}_{c,f}\left[e^{-\beta H_{0}}\mathrm{T}_{\mathcal{C}}\exp\left(\mathcal{S}\left[n_{f}\right]\right)\right]}\\
\mathcal{S}\left[n_{f}\right]= & -i\int_{\mathcal{C}}d\bar{t}\int_{\mathcal{C}}d\bar{t}^{\prime}c^{\dagger}\left(\bar{t}\right)\Lambda\left(\bar{t},\bar{t}^{\prime}\right)c\left(\bar{t}^{\prime}\right)-in_{f}\int_{\mathcal{C}}d\bar{t}U\left(\bar{t}\right)c^{\dagger}\left(\bar{t}\right)c\left(\bar{t}\right)+\left(E_{f}-\mu\right)n_{f}
\end{align}
where $H_{0}=\mu c^{\dagger}c$, $\mathrm{Tr}_{c,f}$ means trace
over $c$ and $f$ degrees of freedom, and $\mathcal{C}$ denotes
Keldysh contour. We introduce the definition 
\begin{equation}
Z_{n_{f}}=\mathrm{Tr}_{c}\left[e^{-\beta H_{0}}\mathrm{T}_{\mathcal{C}}\exp\left(\mathcal{S}\left[n_{f}\right]\right)\right]\label{eq:solver1}
\end{equation}
 with $n_{f}=0,1$. Tracing over the $f$ degrees of freedom, we have
\begin{equation}
G\left(t,t^{\prime}\right)=w_{0}Q\left(t,t^{\prime}\right)+w_{1}R\left(t,t^{\prime}\right)\label{eq:solver}
\end{equation}
where 
\begin{align}
Q\left(t,t^{\prime}\right)= & -i\frac{\mathrm{Tr}_{c}\left[e^{-\beta H_{0}}\mathrm{T}_{\mathcal{C}}\mathcal{S}\left[0\right]c\left(t\right)c^{\dagger}\left(t^{\prime}\right)\right]}{Z_{0}},\\
R\left(t,t^{\prime}\right)= & -i\frac{\mathrm{Tr}_{c}\left[e^{-\beta H_{0}}\mathrm{T}_{\mathcal{C}}\mathcal{S}\left[1\right]c\left(t\right)c^{\dagger}\left(t^{\prime}\right)\right]}{Z_{1}},
\end{align}
and 
\begin{equation}
w_{1}=1-w_{0}=\frac{Z_{1}}{Z_{0}+Z_{1}}\label{eq:solver2}
\end{equation}
 is in fact the definition for $f$-particle density (exactly in the
sense of statistical average). Generally, $w_{1}$ depends on the
$c$-particle configuration and $E_{f}$ by Eqs.~(\ref{eq:solver1})
and~(\ref{eq:solver2}) in a very complicated way and is not known
a priori. We focus on the homogeneous phase of $f$-particles. For
the half-filling we have investigated, we simply have $w_{1}=w_{0}=\frac{1}{2}$. 
\item To obtain $Q\left(t,t^{\prime}\right)$ and $R\left(t,t^{\prime}\right)$,
one can investigate them by equations of motion. Then we transform
all the Green's functions in Eq.~(\ref{eq:solver}) into the Floquet
space. Finally, we obtain Eq.~(10) presented in the manuscript.
\end{enumerate}

\subsection{Charge pump and charge density wave for the equilibrium case}

\begin{figure}[h]
\includegraphics[scale=1.5]{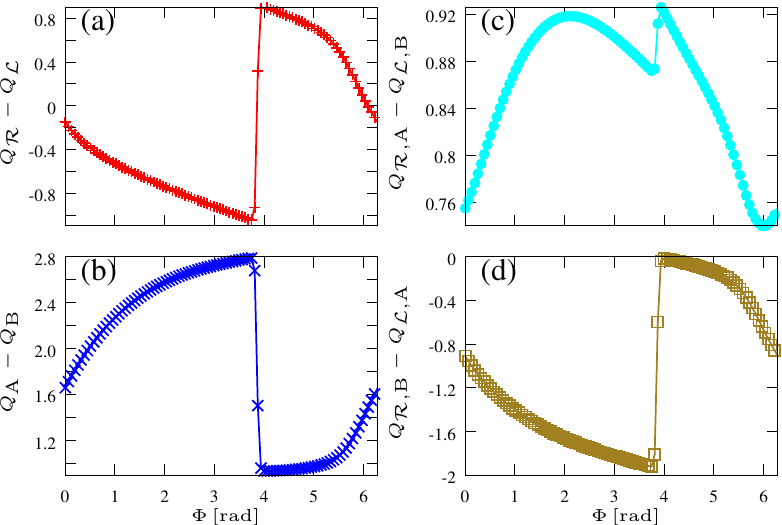}\caption{\label{fig:pump_cdw}Charge pump $Q_{\mathcal{R}}-Q_{\mathcal{L}}$
and charge density wave $Q_{A}-Q_{B}$ for the non-interacting equilibrium
and topological non-trivial case without dissipation. The parameters
are the same as the panel (a) of Fig.~7 in the manuscript.}
\end{figure}

For the equilibrium case, we make a connection between charge pump
and charge density wave (CDW)~\cite{sGrusdt2014rb}. As we argued
in the main text, the difference between charge pump and CDW lies
in the part $Q_{\mathcal{L},\mathrm{A}}-Q_{\mathcal{R},\mathrm{B}}$.
We show in Fig.~\ref{fig:pump_cdw} the charge pump (Fig.~\ref{fig:pump_cdw}.(a))
and CDW (Fig.~\ref{fig:pump_cdw}.(b)) for a non-interacting, dissipationless
and topologically non-trivial case. We observe a jump in both cases,
which signals a possible relationship. For a pronounced integer charge
pump, there must be a dominated process from A to B or from B to A,
which leads to a significant density redistribution between A and
B. As shown in Fig.~\ref{fig:pump_cdw}.(d), the dominated process
is from A to B when $\Phi\approx3.8\mathrm{[rad]}$. It is because
the density on A site is higher than that on B site (Fig.~\ref{fig:pump_cdw}.(b))
due to CDW. Therefore, we can see that a significant CDW favors charge
pump. 

\bibliographystyle{apsrev4-1}

\end{document}